# Foliated Cobordism and Motion

D.H. Delphenich [†]
Physics Department,
University of the Ozarks,
Clarksville, AR, 72840

*Abstract. The mathematical notion of foliated cobordism is presented and its relationship to both the motion of extended particles and waves is detailed. The fact that wave motion, when represented in such a manner on a four-dimensional spacetime, leads to a reduction of the bundle of linear frames to an* SO(2)-*principle sub-bundle is demonstrated. Invariants of foliated cobordism are discussed as they relate to the aforementioned cases of motion.*

[†]   ddelphen@ozarks.edu.

**Contents**

I. INTRODUCTION. Time is perhaps the fundamental enigma of Nature, so it should not be surprising that numerous mathematical techniques exist for modeling the time evolution of states in natural systems. For instance, the simplest is probably that of a single parameter $\tau \in \mathbb{R}$ that defines a curve in the state space whose differentiability class is determined by the nature of the problem. If one wishes to consider objects more extended than points then one can either map the time manifold T into the state space M − as with the flow of a dynamical system − or map the state space into the time manifold; the level surfaces of this (proper) time function $\tau: M \to T$ then become the successive states of the object ([1]).

This last way of looking at time, the parameter, brings us closer to established techniques of differential topology, such as Morse theory and foliations. In the first case, we would be assuming that $\tau$ had non-degenerate critical points, i.e., that wherever $D\tau = 0$, the Hessian of $\tau$, i.e., $D^2\tau$, at that point has non-zero eigenvalues ([2]). In the second case, one generalizes from the level surfaces of a differentiable function to the leaves of a foliation (of codimension one). If the proper-time function has singularities in the form of critical points then one is looking at a singular foliation. The aforementioned example of a dynamical system defines a foliation of dimension one.

---

[1] Indeed, one can also consider the "points" of the curve to live in a – generally infinite-dimensional – space of functions, such as curves, surfaces, wave functions, fields, that define the extended objects. The choice is usually one of PDE's on a finite-dimensional manifold versus ODE's on an infinite-dimensional one.

[2] We shall use the notation D for the differential of a differentiable map, so as not to be confused with *d*, the exterior derivative of a differential form.

Foliations occur naturally in the theory of partial differential equations on manifolds and exterior differential systems by way of the integral manifolds of the system. In particular, the Hamilton-Jacobi equation introduces a notion of complementary foliations: the foliation of codimension one defined by the Hamilton principal function S and the foliation of dimension one that is defined by its transversal trajectories. Together, these complementary foliations define the *Caratheodory complete figure*.

A largely trivial example of a foliation is the product foliation, which is defined by the projection M×N → N, and whose leaves are all of the form M×{y}, hence diffeomorphic. Although the justification for the use of more general foliations than product foliations in the context of physical models generally suggests the existence of singularities of one sort or another – either the kind that keep a foliation from being non-trivial or the kind that show up in the foliation itself – this first examination of the role of foliations and foliated cobordism in various physical models for motion will be concerned with the issues that precede the introduction of singularities, which will be deferred to a later analysis.

The first conclusion being made in this article is that the motion of extended particles in continuum mechanics defines a foliated cobordism of codimension $n-1$, i.e., the cobordant foliations are of dimension zero, and that wave motion defines the more interesting case of codimension one. Furthermore, some of the key theorems of continuum mechanics define integral invariants that are also foliated cobordism invariants.

Since wave motion involves codimension one, the physical significance of the Godbillon-Vey invariant is examined. It is particularly convenient that most of the results that pertain to foliations of codimension one are for three-dimensional manifolds – in part because the Godbillon-Vey form is a 3-form − since they would then relate to the foliation of proper-time simultaneity leaves by the isophase surfaces of a wave.

The second conclusion that we shall derive from this formulation is that when wave motion on a four-dimensional manifold is described by such a foliated cobordism, this necessarily leads to a physically sensible reduction of the bundle of linear frames on spacetime to an SO(2)-principle sub-bundle, i.e., an SO(2)-structure.

The intended audience for this article would consist of theoretical physicists who are reasonably familiar with the geometrical and topological methods of mathematical physics, at least the basic notions of differential manifolds, fiber bundles, and Lie groups. Occasional references to more specialized topics will be made, but can usually be safely ignored as tangential asides without destroying the continuity of the presentation; otherwise, the topic will be discussed *ab initio*. Although some mathematicians might find the applications of the theory of foliations to physical modeling problems amusing, they are forewarned that no attempt has been made here to prove any new mathematical theorems.

**II. FOLIATED MANIFOLDS.** There are various ways of partitioning a manifold into smaller pieces – fibers, orbits, leaves, strata, etc. – and the various ways intersect in various ways as well. The concept that we are going to elaborate upon is that of *leaves*. It generalizes the notion of level surfaces of a submersion $f$: M → N of an *n*-dimensional manifold M into a *k*-dimensional manifold N. This means that D$f$ has rank $k$ at every point of M; a point where this is not true would be called a *singularity* of $f$. For instance, when N = $\mathbb{R}$ the condition that $f$ be a submersion simply means D$f$ ≠ 0 at every point. When $f$ is a submersion, the inverse image of any $y \in$ N will be a submanifold of dimension $n-k$, or *codimension k*. Consequently, these inverse images partition M into lower-dimensional submanifolds. As long as $f$ has no singularities, the submanifolds will be diffeomorphic. The simplest example of a submersion that defines a foliation is the projection M×N → N whose level submanifolds will be all of the M×{$y$}.

The concept of a *foliation* [**33, 40, 52, 54**] of a manifold M is a generalization of this preceding construction. In particular, a foliation of an *n*-dimensional manifold M is a partitioning M = $\vee \Lambda_\alpha$ of M into disjoint submanifolds of dimension $k$ (or codimension $n-k$) − which are called *leaves* – in such a way that M has an "atlas of submersions;" i.e., every $x \in$ M has a neighborhood U and a submersion $\phi$: U → $\mathbb{R}^{n-k}$ in such a way that when two such charts (U, $\phi$) and (V, $\psi$) overlap, there is a diffeomorphism $\Phi_{UV}$ of $\mathbb{R}^{n-k}$ such that on U ∩ V one has: $\psi = \Phi_{UV} \phi$. This has the effect of taking leaves to leaves.

One can weaken this construction by allowing the submersions to have singularities; in that case, one has defined a *singular foliation* [**27**]. For instance, in the case of codimension one, this brings the methods of Morse theory [**37**] into play locally.

Examples of foliations are vast and profound in mathematics. Besides the essentially trivial case mentioned above of foliations by the level submanifolds of a projection, one also has that a fibration defines a foliation of the total space by its fibers, in some instances – such as almost free actions ([3]) [3] – the orbits of a group action can define a foliation, as well as the integral curves of a vector field, and, more generally, the integral submanifolds of a differential system.

The last two examples are the ones that will occupy our attention for the present study. When a vector field X on M is completely integrable, such as when M is compact, it has a global flow whose trajectories (which are also orbits of the action of $\mathbb{R}$) define a foliation of dimension one. If X has no zeroes then one can generalize this example slightly by defining a one-dimensional differential system in the form of a rank-one sub-bundle L(M) of T(M), i.e., a line bundle whose lines are spanned by X. When X has zeroes, its flow has fixed points and one must resort to singular foliations to account for the behavior of its trajectories.

Moreover, when M has a metric (or pseudo-metric), one can decompose T(M) into a Whitney sum:

$$T(M) = L(M) \oplus \Sigma(M).$$

The codimension-one sub-bundle $\Sigma(M)$ that is complementary to L(M) also defines a differential system. Whether or not it defines a complementary foliation of M into submanifolds of codimension one that are transversal to the flow of X is a much deeper issue, which we now address.

---

[3] I.e., ones for which the isotropy subgroup is discrete at any point. This condition is not necessary; consider the natural action of SO(3) on $\mathbb{R}^3 -$ {0}, whose orbits are all 2-spheres and foliate the space even though the isotropy subgroup is SO(2) at every point.

To say that a differential system on M, i.e., a rank $k$ vector sub-bundle V(M) of T(M) is *completely integrable* ([4]) is to say that there is a foliation of M by $k$-dimensional leaves.

The key theorem concerning the complete integrability of differential systems is Frobenius's theorem, which can be stated either in terms of vector fields or 1-forms. Hence, we assume that the fibers of V(M) can also be (locally) expressed as the annihilating subspaces for a set of $n-k$ linearly independent ([5]) 1-forms $\theta^i$. We then have:

**Frobenius's Theorem**:
> *A rank k differential system* V(M) *is completely integrable iff:*
> Whenever X, Y are sections of V(M), [X, Y] is also a section of V(M),
> *or equivalently:*
> For all $i = 1, \ldots, n-k$: $\qquad \theta^i \wedge d\theta^i = 0$.

Since the 3-forms $\theta^i \wedge d\theta^i$ play such an important role, we give them the name of *Frobenius forms* for the differential system; one appears in relativistic hydrodynamics as the Pauli-Lubanski spin vector, which is dual to a related 3-form.

Notice that Frobenius implies that although vector fields do not always have global flows on non-compact manifolds, nevertheless, one-dimensional differential systems are always integrable. The reason that there is no contradiction between their integrability and the local nature of flows is simply due to the fact that a flow implies a specific parameterization of the integral curves, whereas a one-dimensional foliation need only have local parameterizations.

In the case of codimension one, integrability is not guaranteed unless the Frobenius form $\theta \wedge d\theta$ of the differential system vanishes. If it does, then one can also find a 1-form $\eta$ such that $d\theta = \eta \wedge \theta$. From $\eta$, one can form another 3-form, called the *Godbillon-Vey* form, that is associated with the codimension-one foliation $\mathscr{F}$ that is defined by $\theta$:

$$\text{GV}[\mathscr{F}] = \eta \wedge d\eta.$$

It can be shown [**21**] that GV is closed, so it defines an element of $H^3(M, \mathbb{R})$, and even though $\eta$ is not unique, any other choice of $\eta$ would give a cohomologous 3-form. In order for GV to serve as a "characteristic class" for a codimension-one foliation, one needs a notion of equivalence that pertains to foliations and gives us that codimension-one foliations on M will have the same GV class iff they are "equivalent." There are various approaches to this problem, such as integrable homotopy, concordance, and foliated cobordism; we shall mostly discuss the last one first since it pertains to the immediate study.

---

[4] The use of the word "completely" is to emphasize that integral manifolds of dimension less that $k$ may still exist, even if they do not exist for dimension $k$.

[5] Since the global existence of such 1-forms is a matter of vanishing topological obstructions, we shall assume their existence locally.

First, let us briefly digress on the subject of cobordism in its more general context [**50**]. The "general nonsense" of it is that two closed ([6]) $k$-dimensional (…)-manifolds $B_1$ and $B_2$ are said to be (…)-*cobordant* if there is a $k+1$-dimensional (…)-manifold with boundary C such that $\partial C = B_1 \cup B_2$. This defines an equivalence relation on the category of (…)-manifolds and (…)-morphisms. Compactness is essential, since otherwise all closed manifolds B of the same dimension would be the boundary of B×[0, 1), hence, equivalent.

Examples of (…)-manifolds might be: (…) = unoriented, oriented, Lorentz, complex, symplectic, or spin manifolds. We shall first discuss the cases of foliated cobordism and then Lorentz cobordism somewhat later; in particular, foliated cobordism has some idiosyncrasies in the eyes of general nonsense.

Equivalence under cobordism generalizes the situation that exists in the context of the simpler case where, instead of manifolds, one considers $k$-cells in a cellular decomposition of a manifold M. The issue in that event is usually whether a closed $k$-cell is the boundary of some $k+1$-cell. The obstruction to this in the general case is the $k^{th}$ homology module $H_k(M, \mathbb{Z})$; if it does not vanish then some closed $k$-cells are not boundaries. The corresponding ring of (…)-cobordism classes, whose operations are derived from connected sum and intersection, is then a generalized cohomology ring for the category of (…)-manifolds.

One could say that the key issue in the debate over the existence of magnetic monopoles is one of oriented submanifold cobordism: if one is given a closed orientable two-dimensional spacelike submanifold S in spacetime M and a closed 2-form $\Omega$ that represents the electromagnetic field in M then the total magnetic flux through S is $\int_S \Omega$. If S is the boundary of some compact three-dimensional orientable submanifold $S = \partial V$ then, by Stokes's theorem, this total flux should vanish:

$$\int_S \Omega = \int_{\partial V} \Omega = \int_V d\Omega = 0.$$

If S is not such a boundary then the total flux, which is proportional to the magnetic charge contained in S, does not have to vanish, even though $d\Omega$ does. However, since the manifolds we are concerned with are all submanifolds of the same manifold, it is simpler to confine oneself to 2-submanifolds of M that can be decomposed into two-dimensional chain complexes ([7]), so one need only consider the de Rham cohomology ring of M, $H^*(M, \mathbb{R})$, rather than the oriented submanifold cobordism ring of M. In particular, the issue is the vanishing of $H^2(M, \mathbb{R})$.

One would expect that the definition of foliated cobordism would simply involve connecting two foliated manifolds with another one. This is true to some extent, except that one needs to address the way that the leaves are related in all three cases. A first guess might be to make the boundary components $B_1$ and $B_2$ leaves of the foliated manifold C. This would work fine for codimension-one, except that $B_1$ and $B_2$ would not need to be foliated. The definition that seems to be agreed upon by differential topology is that the closed manifolds $B_1$ and $B_2$ with foliations of codimension $k$ are *foliated cobordant* iff there is a foliated manifold C whose foliation also has codimension k, such that $\partial C = B_1 \cup B_2$ and the leaves of C intersect $B_1$ and $B_2$ transversally in leaves of the boundary foliations. The last requirement sounds abstruse but it still keeps us within the purview of physical mechanics. We shall see that both particle motion and wave motion are examples of foliated cobordism; they correspond to codimension $n-1$ and codimension 2, respectively. (In the case of flows, the foliation of the boundary components is the trivial, i.e., zero-dimensional, foliation by points.)

---

[6] I.e., compact, without boundary.

[7] Such as differentiable singular cubic chains.

Something – such as a differential form, an integral, or a group – is said to be a *foliated cobordism invariant* if it is the same for foliated manifolds that are foliated cobordant. GV defines an example of such an invariant, although GV itself is not necessarily a foliated cobordism invariant. Rather, it is the *Godbillon-Vey number* of a closed three-manifold S with a foliation $\mathcal{F}$ of codimension one:

$$\int_S GV[\mathcal{F}],$$

that defines a foliated cobordism invariant.

Actually, since GV is closed, Stokes's theorem implies that it is also an invariant of an oriented cobordism, as well. More generally, Stokes implies that the integral of *any* closed $k$-form $\alpha$ will be an oriented cobordism invariant for $k$-manifolds:

$$\int_{B_2} \alpha - \int_{B_1} \alpha = \int_{B_1 \cup B_2} \alpha = \int_C d\alpha = 0;$$

hence:

$$\int_{B_2} \alpha = \int_{B_1} \alpha.$$

Some particular examples of this situation are the Euler-Poincaré characteristic and the top Pontrjagin number for a $4k$-dimensional manifold, which are obtained by integrating the Euler class and top Pontrjagin class, respectively.

A fundamental result of Thurston [**53**] is that the 3-sphere $S^3$ can be given a continuous infinitude of possible codimension-one foliations that are inequivalent in the eyes of foliated cobordism; they will each have distinct Godbillon-Vey classes and numbers. By comparison, a well-known result of Rohlin [**41**] says that any two compact 3-manifolds are (unoriented) cobordant. Hence, foliated cobordism has a finer degree of resolution as far as the equivalence of manifolds is concerned. We shall return to this result of Thurston's in the context of wave motion.

As we shall see, some of the noted theorems of hydrodynamics take the form of defining foliated cobordism invariants. Moreover, the Godbillon-Vey invariant plays an important role in wave mechanics.

The other concepts that were mentioned above – concordance and integrable homotopy – are progressively more specific forms of foliated cobordism. Two foliations of codimension $k$ on a given manifold M are *concordant* if there is a codimension-$k$ foliation of M×I that is transverse to the "initial and final" boundary components M×{0} and M×{1}. This defines a foliated cobordism in which the initial and final manifolds are diffeomorphic; however, the map that takes leaves of one foliation to leaves of the other might still be non-trivial. In order to make a concordance into an *integrable homotopy,* one must also add the constraint that the leaves of the foliation on M×I must intersect *all* of the intermediate sections M×{$t$} transversally, and not just the initial and final ones.

The last two types of equivalence are especially interesting in the context of integral submanifolds of non-singular differential systems. In order for the cobordism to not take the form of M×I, which makes the initial and final manifolds diffeomorphic, one would have to introduce a singularity, since the initial-value problem involves only flows of diffeomorphisms in the non-singular case.

III. EXTENDED PARTICLE MOTION. The reason that we shall be concerned with extended particles, rather than point particles, is simply that a point particle defines only one curve in a configuration space, whereas to describe a foliation of that space of dimension one we will need a *congruence* of such curves, such as the set of non-intersecting curves that are obtained by the time evolution of each point of an extended object in space. Of course, this also suggests that many of the physical quantities one uses, such as mass density or velocity vector fields, will necessarily have compact spatial support, whether by being spatially bounded in extent or by being asymptotically constant "at infinity." Hence, although spatial compactness is a subtle issue in the cosmological or sub-atomic scale, nevertheless, it is reasonable when you are only using space as a place in which to embed (or immerse, as the case may be) extended objects.

A general picture that emerges out of Hamilton-Jacobi theory, which will be discussed in more detail shortly, is the concept of the *complete figure* for a dynamical system. The terminology is due to Caratheodory in Hamilton-Jacobi theory [**8, 12**], but the concept is more general. In particular, it amounts to a foliated cobordism between codimension-one leaves (geodesically equidistant hypersurfaces) in a foliated manifold M which are themselves foliated by points (leaves of codimension-$n$-1 in the boundary components) and are connected by leaves of codimension-$n$-1 in M that are transversal to the foliations of the boundary components, i.e., the integral curves of the flow of the dynamical system. We will now examine various mathematical and physical contexts in which the general notion of a foliated cobordism of this nature arises.

A. *Gradient dynamical systems.* When M has a metric or pseudo-metric $g$ so we can define the 1-form $\theta = i_X g$ then in the event that $d\theta = 0$, we can define the system of first-order PDE's:

$$\theta = d\phi.$$

If this is the case, X is said to be a *gradient dynamical system* since that would make:

$$\theta = i_X g = d\phi, \quad \text{i.e.,} \quad X = \nabla \phi, \qquad \text{for some } \phi \in C^\infty(M).$$

By the Poincaré lemma, this system has local solutions, but their global existence depends on the vanishing of $H^1(M, \mathbb{R})$, the first de Rham cohomology group of M, or, by the Hurewicz isomorphism theorem, its fundamental group $\pi_1(M)$.

Gradient dynamical systems lead naturally to the methods of Morse theory ([8]). The zeroes of X, which represent fixed points of its flow, are also critical points of the potential function $\phi$. The stability matrix of X, namely, the matrix of DX, at a fixed point then becomes $D^2\phi$, the Hessian of $\phi$, at that point, and one sees that when the Hessian is non-degenerate there is a close relationship between the stability type of the zeroes of X, i.e., the eigenvalues of DX, and the topological type of the submanifolds $\phi^{-1}(\tau \leq \tau_0)$ of M that are bounded by the level surfaces of $\phi$; in particular, the hyperbolic fixed points of X will be associated with the attachment of *k*-cells, where *k* is the number of negative eigenvalues for $D^2\phi$.

---

[8] More generally, one considers *Morse-Smale* dynamical systems.

In addition to the one-dimensional foliation by integral curves, gradient dynamical systems also define a (possibly singular) codimension-one foliation of M by the level surfaces of $\phi$. Since $X = \nabla\phi$ this makes the one-dimensional foliation by integral curves of X transversal to the level surfaces of $\phi$, except possibly in the pseudo-Riemannian case, where X might be light-like, but orthogonal to light-like level surfaces. Hence, in the "non-characteristic" case, we have a simple example of transverse foliations of the same codimension (namely, $n-1$): the foliation by level surfaces of $\phi$, which are themselves foliated by points, and the foliation by integral curves. Hence, this is an elementary example of foliated cobordism.

In the case of the electric field strength vector field **E** in electrostatics, the flow of **E** defines the field lines, and its transversal hypersurfaces are the equipotentials. However, in the electrodynamic case, for which $\nabla\times\mathbf{E}$ might not be zero, such equipotentials cannot exist. An analogous situation occurs in hydrodynamics, where the vector field in question is the velocity vector field **v**, and the curl represents its vorticity; if the vorticity is non-vanishing, one cannot have a potential flow.

There is another way of (possibly) foliating M by using X: the level surfaces of a constant of the motion *f*. A constant of the motion defined by X is a smooth function *f* on M such that the values of *f* are constant along the integral curves of X, i.e.,

$$L_X f = Xf = 0.$$

Since this equation can also be given the form $g(X, \nabla f) = 0$, one sees that these level surfaces contain the integral curves of X and are orthogonal to the level surfaces of $\phi$. Hence, this gives us another way of defining a foliated cobordism between the leaves defined by $\phi$: we give the level surfaces of a codimension-one foliation by their (transverse) intersections with the level surfaces of *f* and connect the equipotentials by those level surfaces. Note that if we find *two* constants of motion $f_1$ and $f_2$ that are independent, in the sense that the map $f_1\times f_2: M \to \mathbb{R}^2$, $x \mapsto (f_1(x), f_2(x))$ has maximal rank then we can define a foliated cobordism of codimension two. Proceeding in this manner, one eventually hopes to reach a point where there are $n-1$ independent constants of motion, in which case, the map $f_1\times f_2\ldots\times f_{n-1}: M \to \mathbb{R}^{n-1}$ should define a coordinate chart on each level surface of $\phi$. This is the general objective of Hamilton-Jacobi theory, which we shall also discuss later.

If one defines a local coordinate system (U, $x^i$) with $i = 1, 2, \ldots, n$ on an open subset $U \subset M$ and expresses the vector field in the form:

$$X = X^i \frac{\partial}{\partial x^i},$$

then the system of ODE's takes the form:

$$\frac{dx^i}{d\tau} = X^i,$$

the PDE that defines the first integrals of X takes the form:

$$Xf = X^i \frac{\partial f}{\partial x^i} = 0,$$

and the system of *n* PDE's that defines the transversal hypersurfaces to X takes the form:

$$X^i = g^{ij} \frac{\partial \phi}{\partial x^j}.$$

Since we are assuming that M has a Riemannian or pseudo-Riemannian metric $g$ we may define two more notions in terms of X: the dual 1-form:

$$\theta = i_X g$$

and the complementary sub-bundle $\Sigma(M)$ to the line bundle $L(M)$ that is spanned by X in $T(M)$:

$$T(M) = L(M) \oplus \Sigma(M).$$

One can also regard $\Sigma(M)$ as consisting of the annihilating subspaces to $\theta$:

$$\theta(\mathbf{v}) = 0, \qquad \text{for all } \mathbf{v} \in \Sigma(M).$$

Although $L(X)$ is always integrable, the integrability of $\Sigma(M)$ is not as necessarily implicit, and is equivalent to the vanishing of $\theta \wedge d\theta$. The leaves of that foliation would then represent "sections" of the flow of $\mathbf{v}$.

B. *Hamiltonian vector fields.* A situation that is similar to the gradient flow is found on symplectic manifolds [**11, 20, 42, 48**], i.e., a manifold M on which one has a closed non-degenerate 2-form $\Omega$. The non-degeneracy allows one to define a linear isomorphism of each $T_x(M)$ and $T_x^*(M)$ by way of $\mathbf{v} \mapsto i_\mathbf{v}\Omega$. Although this is analogous to what a metric would do on M, there are two important differences. For one thing, although examples non-isometric metrics on M are commonplace, there is only one equivalence class of symplectic structure, up to symplectomorphism, i.e., up to diffeomorphisms that preserve the symplectic 2-forms; this is the essential content of Darboux's theorem. Also, the symplectic gradient vector field of a smooth function (which we shall define shortly) is not transverse to the level surfaces of that function, but tangent to them; this is referred to as *skew-orthogonality*.

A vector field X on a symplectic manifold M is called *globally Hamiltonian* ([9]) if there is an $H \in C^\infty(M)$ such that X is the *symplectic gradient* of H, i.e.:

$$i_X \Omega = dH, \qquad \text{or} \qquad X = \nabla_\Omega H.$$

One justification for the terminology is that if we choose a *canonical coordinate system* $\{U, (x^i, p^i)\}$, i.e., one whose existence is demanded by Darboux and makes the symplectic 2-form take the local expression:

$$\Omega = dp^i \wedge dx^i = \tfrac{1}{2} J_{ij}\, dx^i \wedge dp^j, \text{ where } J_{ij} = \begin{bmatrix} 0 & -I \\ I & 0 \end{bmatrix}.$$

then, relative to the natural frame of this coordinate system, a globally Hamiltonian vector field $X_H$ will have the component expression:

$$X_H = \nabla_\Omega H = J^{-1} dH = \frac{\partial H}{\partial p^i} \frac{\partial}{\partial x^i} - \frac{\partial H}{\partial x^i} \frac{\partial}{\partial p^i},$$

for some smooth function $H \in C^\infty(M)$.

Hence, the system of ODE's that $X_H$ defines $\dot{\gamma} = \nabla_\Omega H$ takes the local form:

---

[9] A *locally Hamiltonian vector field* X is one for which $i_X \Omega$ is closed, but not exact; for simply connected manifolds, this is equivalent to the definition of a globally Hamiltonian vector field.

$$\frac{dx^i}{dt} = \frac{\partial H}{\partial p^i}, \qquad \frac{dp^i}{dt} = -\frac{\partial H}{\partial x^i},$$

which one recognizes as Hamilton's equations.

The first PDE that $X_H$ defines $X_H f = 0$ is the equation of the functions that are constants of the motion, or *first integrals*. Due to the special nature of $X_H$, this equation takes the form:

$$X_H f = \Omega(\nabla_\Omega H, \nabla_\Omega f) = -\{H, f\} = 0.$$

That this notation is a learned borrowing from the classical *Poisson bracket* of H and $f$ follows from the fact that in a canonical coordinate system the general expression becomes:

$$\{f, g\} = -\Omega(\nabla_\Omega f, \nabla_\Omega g) = \frac{\partial f}{\partial x^i}\frac{\partial g}{\partial p^i} - \frac{\partial g}{\partial x^i}\frac{\partial f}{\partial p^i}.$$

In particular, note that H is constant along the integral curves of $X_H$. This can be given the usual interpretation of energy conservation if H represents the total energy density of a mechanical system.

The codimension-one foliation of M that is *symplectic-dual* to the integral curves is defined by the level surfaces of H, i.e., the constant-energy hypersurfaces. However, since H is constant along the integral curves of $\nabla_\Omega H$, these integral curves must each lie *within* a corresponding leaf of this codimension-one foliation. Hence, the two foliations are not transversal, so if we want to find a complete figure that is defined by $\nabla_\Omega H$ we will have to look further.

C. *Hamilton-Jacobi theory*. In the previous section, we were concerned only with motion in the *phase space* of a Hamiltonian system, i.e., the symplectic manifold in question. Now we shall direct our attention to the motion of our system in its configuration manifold.

A common example of a symplectic manifold that is of interest to mechanics is the cotangent bundle T*(M) for any configuration manifold, M. The symplectic 2-form is defined as $\Omega = d\theta$, where the canonical 1-form $\theta$ is defined by:

$$\theta_p(\mathbf{v}) = p(\pi_* \mathbf{v}),$$

whenever $\mathbf{v} \in T_p(T_x^*(M))$. In a canonical coordinate system this 1-form is simply:

$$\theta = p_i dx^i.$$

This 1-form on T*(M) should not be confused with a 1-form on M, i.e., a section $p: M \to T^*(M)$, even though they would have similar local expressions:

$$p = p_i(x) dx^i.$$

The difference is that the $p_i$'s that appear in $\theta$ are functions on T*(M), not functions on M. However, for any section, $p$, we will have: $p^*\theta = p$. Our use of the notation, $p$, is suggestive of the fact that sections of T*(M) can be interpreted as momentum (co-velocity, resp.) 1-forms that are associated with velocity vector fields by way of the energy-momentum tensor (metric tensor, resp.).

Now that we know how $\theta$ pulls down along $p$, we also need to know how $\Omega$ pulls down along $p$. A particular case of interest is the case where $p$ is a *geodesic section* ([10]), i.e.:

$$p^*\Omega = 0.$$

The injective differentiable map $p: M \to T^*(M)$ defines a submanifold $(p, M)$ of $T^*(M)$, but it does not have to be a diffeomorphism onto, or even an immersion, since $dp$ can have zeroes. The fact that $p^*\Omega$ vanishes is also expressed by saying that the submanifold $(p, M)$ is *isotropic*. Since the dimension $n$ of this submanifold is maximal for an isotropic submanifold, it is, by definition, a *Lagrangian submanifold* of $T^*(M)$. The more general theory of Lagrangian submanifolds of symplectic manifolds is also concerned with submanifolds that do not take the form of geodesic sections of the cotangent bundle, such as immersed submanifolds of $T^*(M)$, for which there might be projective singularities, such as folds and cusps.

From the fact that $d$ commutes with pull-back, we then see that for a geodesic section of $T^*(M)$, we have:

$$0 = p^*d\theta = d(p^*\theta) = dp,$$

i.e.:

$$dp = 0.$$

Mathematically, this says that $p$ must be closed; physically, if $p$ is the momentum 1-form for the flow of some extended object then that flow must have vanishing dynamical vorticity. Note that this is also a sufficient condition for the integrability of the differential system defined by the annihilating subspaces of $p$.

A stronger condition on $p$ than being closed would be to demand that $p$ be exact, so:

$$p = dS$$

for some 0-form S which is called *Hamilton's principal function*. (Of course, if M is simply-connected, so that $H^1(M, \mathbb{R}) = 0$, this condition is equivalent to being closed; in particular, this is always true locally.) Now suppose that $p$ takes its values in a level surface of $H \in C^\infty(T^*(M))$:

$$p^*H = H \circ p = E,$$

where E is a constant ([11]).

The pair of equations that we have then defined:

$$\begin{cases} p = dS, \\ H \circ p = E \end{cases}$$

define the time-invariant form of the *Hamilton-Jacobi* equation.

When combined and viewed locally, these equations take on the more conventional form:

$$H(x^i, \frac{\partial S}{\partial x^i}) = E.$$

We can express this picture in the language of foliations. The function S defines a (possibly singular) codimension-one foliation of M into geodesically equidistant hypersurfaces, just as H defines a codimension-one foliation of T*(M) into constant energy hypersurfaces. We can then say that the geodesic field $p$ maps the foliation of M defined by S into a *single* leaf of the one defined on T*(M) by H. If we turn the 1-form $p$ into a vector field **v** by way of a metric or energy-momentum tensor then the integral curves of **v** are transverse to the leaves defined by S. Hence, they define a codimension-$n-1$ foliated cobordism between those leaves, which gets mapped into $H^{-1}(E)$ by $p$.

---

[10] The use of the word "geodesic" will be justified shortly when we discuss geodesic flows; for now, let it suffice to say that it is also used more generally in the usual Hamilton-Jacobi theory [15].

[11] When E = 1, the level surface $H^{-1}(1)$ is called the *figuratrix* of the system.

Two ways of generalizing $p$ without leaving behind codimension-one foliations are $p$ closed, but not exact, and $p$ integrable, but not closed. The first case, which is only possible when M is not simply connected, implies that the foliation is only locally representable by level surfaces of a smooth function. The second case implies that the $p \wedge dp$ vanishes, even though $dp$ does not. In this case, the codimension-one foliation is locally representable by level surfaces of some function $f$ but $df$ is not even locally equally to $p$. Furthermore, one could allow $p$ to have singularities (i.e., zeroes, fixed points, stagnation points, etc.); in those cases, the foliations would be singular, as well

To get the full time-varying form of the Hamilton-Jacobi equation, which one needs for non-conservative systems, such as dissipative systems or systems not in equilibrium, one needs to extend from a symplectic structure to a *contact structure.* The difference is mainly one of using a closed 2-form $\Omega$ of rank one instead of rank zero ([12]). This implies that a (finite-dimensional) contact manifold must be odd-dimensional. If $\mathbb{R}$ represents the proper time manifold then we can form such a contact manifold from $J^1(\mathbb{R}, M)$, the bundle of 1-jets of differentiable curves in M [**45**]. The 1-*jet* of a curve $\gamma: \mathbb{R} \to M$ at a point $x \in M$, which is notated by $j_x^1 \gamma$, is the equivalence class of all differentiable curves through $x$ that have the same tangent at $x$; this is also one way of characterizing a tangent vector at $x$. There is nevertheless a subtle difference between $J^1(\mathbb{R}, M)$ and T(M), since $J^1(\mathbb{R}, M)$ is naturally fibered over $\mathbb{R}$, as well as M. Moreover, $J^1(\mathbb{R}, M)$ is of dimension one higher than T(M); however, there is also a natural fibration $J^1(\mathbb{R}, M) \to T(M)$ that takes any jet to the corresponding tangent vector.

Locally, $J^1(\mathbb{R}, M)$ looks like $(\tau, x^i, p_i)$, so locally, it looks like $\mathbb{R} \times T^*(M)$. If $d\tau$ is the canonical 1-form on $\mathbb{R}$, and H is our Hamiltonian function on $J^1(\mathbb{R}, M)$ then we can define the *Poincaré-Cartan* form [**9, 14, 21, 22, 24, 25, 45, 48**]:

$$\mathcal{L} = \theta - H d\tau.$$

Our use of the notation $\mathcal{L}$ is suggestive of the fact that when this 1-form is pulled down by $p$ and integrated along a curve in M, the resulting number is the action along that curve as defined by H. Once again, if we pull $\mathcal{L}$ down by a geodesic field $p$, and assume that the result is exact:

$$p^*\mathcal{L} = p - (H \circ p)\, d\tau = dS,$$

then the resulting equation:

$$p - (H \circ p)\, d\tau = dS,$$

has the local form of the conventional time-varying Hamilton-Jacobi equation:

---

[12] By the *rank* of a differential form at a point, we mean the maximal dimension of its annihilating subspaces at that point; if this number is constant for all points then we speak of the rank of the differential form itself.

$$\begin{cases} \dfrac{\partial S}{\partial x^i} = p_i, \\ \dfrac{\partial S}{\partial \tau} = -H(\tau, x^i, p_i). \end{cases}$$

In the time-varying case, $p$ would define a *Legendrian submanifold* of the contact manifold $J^1(\mathbb{R}, M)$. From the perspective of Lagrangian (Legendrian, resp.) submanifolds, the time evolution of Lagrangian (Legendrian, resp.) submanifolds by symplectomorphisms (contactomorphisms, resp.) – i.e., the flows of Hamiltonian vector fields – then represents a generalization of the time evolution of geodesic vector fields that are subject to the Hamilton-Jacobi equation.

D. *Geodesic flows.* When M has a (pseudo-)metric $g$ that is defined on T*(M), a natural choice of Hamiltonian might be:

$$H: T^*(M) \to \mathbb{R}, \qquad p_x \mapsto g_x(p_x, p_x).$$

This choice is closely related to kinetic energy – at least for point particles whose mass is constant. As such, it will be sensitive to the choice of parameterization for a given integral curve (i.e., a "faster" curve will have a higher kinetic energy).

Another possible choice of Hamiltonian is:

$$H: T^*(M) \to \mathbb{R}, \qquad p_x \mapsto \sqrt{g_x(p_x, p_x)}.$$

This choice corresponds to the arc-length along the integral curve. By comparison to the previous choice, this choice is independent of the parameterization for the integral curve.

In a canonical coordinate system, the Hamiltonian vector field $\nabla_\Omega H$ that is associated with this latter H has the local form:

$$\nabla_\Omega H = \left(\frac{1}{H} g^{ij} p_j\right) \frac{\partial}{\partial x^i} - \left(\frac{1}{2}\frac{1}{H}\frac{\partial g^{ij}}{\partial x^k} p_i p_j\right) \frac{\partial}{\partial p_k}.$$

Its flow is called the *geodesic spray* for $g$. This represents the result of one integration of the equations of motion on T*(M). A second integration, in the form of choosing a section $p$ of the fibration, e.g., choosing initial conditions, produces a flow on M that is obtained by projection of the geodesic spray along the section $p$. This flow on M is called the *geodesic flow* of $g$. In particular, if $\Phi_\tau: U \to T^*(M)$ is a local flow for $\nabla_\Omega H$ on T*(M) and $p$ is a section, then the local geodesic flow for $\nabla_\Omega H$ along $p$ is:

$$\pi \circ \Phi_\tau \circ p: \pi(U) \to M.$$

When the integral curves are assumed to have unit-speed, so $H = 1$, Hamilton's equations for this H are:

$$\begin{cases} \dfrac{dx^i}{d\tau} = g^{ij} p_j \\ \dfrac{dp_k}{d\tau} = -\tfrac{1}{2} \dfrac{\partial g^{ij}}{\partial x^k} p_i p_j \end{cases}$$

The first equation identifies $p_j$ as the components of the co-velocity 1-form for the flow, i.e., its momentum 1-form with unit mass. The second one can be reorganized into the form of the geodesic equation for the 1-form $p$:
$$\nabla_{\mathbf{v}} p = i_{\mathbf{v}}[(dp_i - \varpi_i^j p_j)dx^i] = 0,$$
using the Levi-Civita connection on T*(M) that $g$ defines:
$$\varpi_j^i = \tfrac{1}{2}[g^{ik}(g_{jm,k} + g_{mk,j} - g_{jk,m})]dx^m.$$
This makes $p$ a geodesic field in the metric sense.

The natural question to ask then is that of how the usage of the term "geodesic field" in the general Hamilton-Jacobi sense relates to its present use in the metric sense. We phrase our answer as a:

**Theorem:**
> *When* H *is the Hamiltonian on* T*(M) *that is defined by arc-length relative to a metric, a section p of that bundle is geodesic in the geometric sense iff it is geodesic in the Hamilton-Jacobi sense.*

**Proof:**
Assume we are dealing with such an H. Observe that for each $x \in M$ the integral curves of $\nabla_\Omega H$ in T*(M) represent the various geodesics through $x$ for the various choices of co-velocity $p_x$. An arbitrary section $p$ of T*(M) will not necessarily take the integral curves of its corresponding velocity vector field to integral curves of the geodesic spray; i.e., the co-velocity 1-form will not necessarily be parallel-translated along the flow. Hence, to say that $p$ is the co-velocity of a geodesic velocity vector field $\mathbf{v}$, i.e., $\nabla_{\mathbf{v}} \mathbf{v} = 0$ is to say that it maps the leaves of the one-dimensional foliation of M by integral curves of $\mathbf{v}$ into leaves of the one-dimensional foliation of T*(M) by integral curves of $\nabla_\Omega H$. But since these respective integral curves are tangent to the respective vector fields $\mathbf{v}$ and $\nabla_\Omega H$ this says that D$p$ must embed $\mathbf{v}$ in $\nabla_\Omega H$, or dually, that it must pull back $\mathcal{L}$ to $d$S. Hence, $p$ is a geodesic section in the metric sense iff it is a geodesic section in the Hamilton-Jacobi sense.

E. *First-order partial differential equations.*  One can generalize beyond the aforementioned methods applied to the Hamilton-Jacobi equation to the class of all first-order partial differential equations in the functions defined on a given manifold. The main difference is that the contact manifold that one works with is $J^1(M, \mathbb{R})$, the bundle of 1-jets of differentiable functions on M, not $J^1(\mathbb{R}, M)$. As a result, the vector fields that correspond to a function on the space that we shall define are the characteristic vector fields that define the characteristic equations of the first-order PDE in question. In effect, the method of characteristics establishes a link between the contact structure of $J^1(\mathbb{R}, M)$ and that of $J^1(M, \mathbb{R})$. It essentially amounts to the difference between regarding proper time as a curve parameter and regarding it as a smooth function on M. We now make these notions precise.

A *first-order partial differential equation* [**1, 8, 10, 11, 12, 15, 24, 30, 42**] is a level surface of a differentiable function:
$$F: J^1(M, \mathbb{R}) \to \mathbb{R};$$

customarily, it is the level surface of 0.

Locally, i.e., when one chooses a coordinate chart (U, $x^i$) on M, the corresponding neighborhood in $J^1(M,\mathbb{R})$ has the form:

$$(x^i, u, p_i), \qquad i = 1, \ldots, n.$$

The coordinates $p_i$ parameterize the possible values of the first derivatives of $u$ at $x$. Hence, the local form of the PDE defined by F would be:

$$F(x^i, u, p_i) = 0.$$

Of course, it is only for the points of this level surface for which $(x_i, u)$ is the graph of a function $u$ on U and $p_i = \dfrac{\partial u}{\partial x^i}$ that we are actually defining a PDE in the conventional sense. Any $C^1$ function $f$ on M defines an $n$-dimensional submanifold $j^1f$: M $\to J^1(M,\mathbb{R})$, which is called the *1-jet prolongation of f,* by way of the set of all 1-jets of $f$ at every point of M. Locally, these points will have the form:

$$(x^i, f(x^i), \left.\dfrac{\partial f}{\partial x^i}\right|_{x^i}).$$

Hence, to say that a function $f$ is a *solution* to the PDE defined by F is to say that:

$$F(j^1f) = 0.$$

Of course, $j^1f$ does not exhaust the entire level surface $F^{-1}(0)$. This is why we define the *contact structure* on $J^1(M,\mathbb{R})$ by the 2-form $d\theta$ where the canonical 1-form $\theta$ on $J^1(M,\mathbb{R})$ takes the local form:

$$\theta = du - p_i dx^i.$$

It has the basic property that it vanishes at a point $j^1$ of $J^1(M,\mathbb{R})$ iff $j^1 = j^1f$ for some $f$ on M. Hence, it picks out the 1-jet prolongations of functions from the arbitrary 1-jets. Therefore, if $f$ is a solution of the PDE defined by F, as above, then one must have:

$$\left.\theta\right|_{j^1f} = 0.$$

The 1-form $\theta$, which is everywhere non-zero, defines a $2n$-dimensional sub-bundle of $T(J^1(M,\mathbb{R}))$ by its annihilating subspaces. The Frobenius 3-form associated with $\theta$ is:

$$\theta \wedge d\theta = (du - p_i dx^i) \wedge dx^i \wedge dp_i = du \wedge dx^i \wedge dp_i.$$

However, this form does not vanish for all jets, although it vanishes for 1-jet prolongations; thus, $\theta$ does not foliate $J^1(M,\mathbb{R})$.

If we were dealing with a first-order ODE instead of a first-order PDE then we would know that there is more than one element to the space of solutions and that to define a unique element would require specifying the value of the solution at one point; for instance, one defines an *initial-value* problem. For a first-order PDE, the situation is somewhat more elaborate, because one then defines the corresponding initial-value problem – the so-called *Cauchy problem* – by specifying the value of $f$ on some $n$-1-dimensional submanifold S of M. If one can define a "characteristic vector field" in terms of F that specifies the "time evolution" of each individual point of S then the Cauchy problem for a first-order PDE amounts to an infinitude of initial-value problems for the points of S subject to an ODE.

Given the aforementioned local form for $\theta$, the contact form becomes:

$$\Omega = d\theta = dx^i \wedge dp_i,$$

which generalizes the symplectic form on T*(M). In the present case, it is no longer non-degenerate, but admits a non-zero vector field X on $J^1(M, \mathbb{R})$ that has the property:

$$i_X \Omega = 0.$$

Because exterior differentiation is a local operator, we also have:

$$\Omega|_{j^1 f} = 0.$$

This is expressed by saying that the submanifold $j^1 f$ is *isotropic* in the contact manifold $J^1(M, \mathbb{R})$. Since it is also *n*-dimensional, which is the maximum possible dimension for an isotropic submanifold of a 2*n*+1-dimensional contact manifold, one sees that every 1-jet prolongation – *a fortiori,* every solution of a first-order PDE on M – defines a Legendrian submanifold of $J^1(M, \mathbb{R})$.

The function F that defines the PDE can be treated like a "generalized Hamiltonian" on the contact manifold in question. The 1-form $d$F defines a vector field X on $J^1(M, \mathbb{R})$ when it is restricted to any submanifold on which $\Omega$ is non-degenerate, such as $j^1 f$:

$$i_X \Omega = d\mathrm{F}|_{j^1 f}.$$

The intersection of a tangent space to the level surface of 0, i.e., $d$F(**v**) = 0, with the annihilating subspace for $\theta$, is a unique line, which defines the *characteristic line field* of the PDE defined by F. It is generally spanned by some non-zero vector field X on $J^1(M, \mathbb{R})$ that is defined up to a non-zero constant at each point.

In local form:

$$X = X^i \frac{\partial}{\partial x^i} + U \frac{\partial}{\partial u} + P_i \frac{\partial}{\partial p_i}$$

$$d\mathrm{F} = \mathrm{F}_{x^i} dx^i + \mathrm{F}_u du + \mathrm{F}_{p_i} dp_i.$$

When $d$F is restricted to $j^1 f$, so that $du = p_i dx^i$, it becomes:

$$d\mathrm{F}|_{j^1 f} = (\mathrm{F}_{x^i} + p_i \mathrm{F}_u) dx^i + \mathrm{F}_{p_i} dp_i.$$

Meanwhile:

$$i_X \Omega = i_X(dp_i \wedge dx^i) = X^i dp_i - P_i dx^i.$$

Hence, by equating components and taking into account that $U = p_i X^i$, since we are considering a 1-jet prolongation of a function, we get the component form of the characteristic vector field for F:

$$X = \mathrm{F}_{p_i} \frac{\partial}{\partial x^i} + p_i \mathrm{F}_{p_i} \frac{\partial}{\partial u} - (\mathrm{F}_{x^i} + p_i \mathrm{F}_u) \frac{\partial}{\partial p_i}.$$

Its integral curves are the *characteristics* of the PDE defined by F. They will be solutions of the *characteristic equations:*

$$\begin{cases} \dot{x}^i = \dfrac{\partial F}{\partial p_i} \\ \dot{u} = p_i \dfrac{\partial F}{\partial p_i} \\ \dot{p}_i = -\left(\dfrac{\partial F}{\partial x^i} + p_i \dfrac{\partial F}{\partial u}\right). \end{cases}$$

When F does not depend on $u$, we can project $J^1(M,\mathbb{R})$ onto $T^*(M)$ so that the contact structure on the former projects onto the symplectic structure of the latter, and if F plays the role of Hamiltonian, we see that the characteristic equations for the (time-invariant) Hamilton-Jacobi PDE are Hamilton's ODE's. However, not every first-order PDE on $J^1(M,\mathbb{R})$ takes the form of a (time-varying) Hamilton-Jacobi equation for some Hamiltonian H on $J^1(\mathbb{R},M)$.

A more general method for finding the characteristic line field, which we will use again in the context of wave motion is the method of characteristic systems of exterior differential systems that was described by Cartan [**7, 10, 11, 20, 24, 30**].

First one represents the first-order PDE by the vanishing of the pair of differential forms F and $\theta$ on $J^1(M,\mathbb{R})$:

$$0 = F$$
$$0 = \theta.$$

The set $\{F, \theta\}$ generates an ideal $I\{F, \theta\}$ in the exterior algebra $\Lambda^*(J^1(M,\mathbb{R}))$, which one regards as the *exterior differential system* defined by F and $\theta$. A typical element $\alpha$ of this ideal takes the form:

$$\alpha = \beta \wedge \theta, \qquad \text{for some } \beta \in \Lambda^*(J^1(M,\mathbb{R})).$$

The Frobenius condition for integrability of $I\{F, \theta\}$ is that it be closed, i.e., $dI \subset I$. Hence, if we want integrability, we need to add $dF$, $d\theta$ to the generators:

$$0 = F \qquad 0 = dF$$
$$0 = \theta \qquad 0 = d\theta.$$

The *characteristic system* of the ideal $I\{F, \theta\}$ is the *associated system* of its *closure* $I\{F, dF, \theta, d\theta\}$. The associated system, in turn, consists of the set of 1-forms on $J^1(M,\mathbb{R})$ that generate $I\{F, dF, \theta, d\theta\}$. To find this system, we look for all vector fields X on $J^1(M,\mathbb{R})$ that take the generators of $I\{F, dF, \theta, d\theta\}$ to other elements of $I\{F, dF, \theta, d\theta\}$:

$$0 = i_X dF = F_i X_i + F_u X_u + F_{p_i} X_{p_i}$$
$$0 = i_X \theta = X_u - p_i X_i$$
$$\phantom{0 = } i_X d\theta = -X_{p_i} dx^i + X_i dp_i$$
$$\phantom{0 = i_X d\theta} = \alpha dF + \beta \theta, \qquad \text{for some } \alpha, \beta.$$

From the last equation, we get $\beta = -\alpha F_u$, and by substitution we get the same characteristic vector field as before, up to multiplication by $\alpha$, which does not affect the foliation by characteristic curves.

The basic method for solving the Cauchy problem for a first order PDE by the method of characteristics is:

1) Map the initial (Cauchy) data $(S, \varphi)$ into $J^1(M, \mathbb{R})$ by means of 1-jet prolongation.
2) (When possible) propagate each prolonged 1-jet along the integral curve of the characteristic vector field X on $J^1(M, \mathbb{R})$.
3) Project the propagated 1-jets back onto the graph of a function:
$$J^1(M, \mathbb{R}) \to M \times \mathbb{R}, \quad j^1 f_x \mapsto (x, f(x)).$$

The caveat in 2) is simply that this process only works when the Cauchy datum $(x, \varphi(x))$ is *non-characteristic,* i.e., the projection of X at $j^1\varphi_x$ is a tangent vector in $T_xM$ that is transversal to S. This also amounts to saying that the projections of the characteristic curves in $J^1(M, \mathbb{R})$ intersect S transversally. Locally, this amounts to saying that the projection of $F_{p_i}$ is transversal to $T_xS$. The foregoing picture is schematically depicted in Fig. 1.

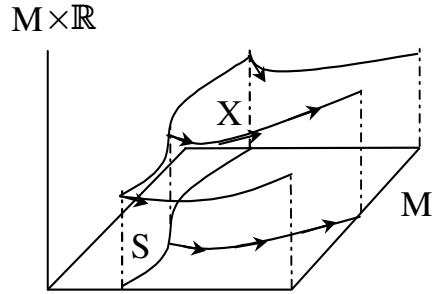

Fig. 1. Solution of the Cauchy problem by characteristics.

This is evidently the same situation we discussed previously for the Hamilton-Jacobi equation, and we summarize the analogies between the two pictures in Table I.

Table I. Analogies between Hamiltonian mechanics and First order PDE's.

|  | Hamiltonian | First Order PDE |
| --- | --- | --- |
| Jet space | $J^1(\mathbb{R}, M)$ | $J^1(M, \mathbb{R})$ |
| Canonical 1-form | $p_i dx^i$ | $du - p_i dx^i$ |
| Defining function | H | F |
| Level surface | Energy | PDE |
| Solutions | Curve in M | Function on M |
| Isotropic submanifold | Geodesic field | 1-jet prolongation of solution |
| Characteristic equations | Hamilton's equations | Characteristic vector field |

Consequently, we see that if we look at an initial Cauchy manifold $S_0$ and its propagated image, or *Cauchy development,* at a later "time point ([13])" $S_\tau$, then the congruence of projected characteristic curves that intersect both $S_0$ and $S_\tau$ define a (possibly singular) foliated cobordism between the successive evolutes of the initial manifold. By prolongation, this cobordism gets mapped into the 0-leaf of the codimension-one foliation on $J^1(M,\mathbb{R})$ that is defined by F.

The possibility of singularities in the foliation relates to the projection $J^1(M,\mathbb{R}) \to M\times\mathbb{R}$. Conceivably, even a non-singular flow in $J^1(M,\mathbb{R})$ might develop folds, cusps and the like under projection back onto $M\times\mathbb{R}$. We shall return to the question of singularities later.

F. *Proper time foliations*: So far, we have been largely concerned with non-relativistic motion, for which there is an implicit assumption that the simultaneity of physical events is well defined. We have, moreover, assumed that simultaneity takes the form of the codimension-one foliation of our Galilean space-time $\mathbb{R}\times\Sigma$ that is defined by projection on the second factor. This has the effect of equating the proper time parameter with the time dimension.

In the relativistic context, simultaneity in the eyes of proper time is no longer consistent for different observers, even for the special relativistic arena of Minkowski space. Furthermore, if we look at the context of general relativity, in which Minkowski space gets replaced with a Lorentz manifold M even the issue of whether M can be expressed as $\mathbb{R}\times\Sigma$ at all may be topologically subtle. We now examine some of these issues in more detail.

A manifold M is said to admit a *Lorentz structure* [23, 36, 49] iff any of the following equivalent conditions are satisfied:
1) M admits a pseudo-metric *g* of globally hyperbolic normal type.
2) M admits a global line field L(M), i.e., a line bundle.
3) M admits a global section to its projective tangent bundle $\mathbb{R}T(M)$.
4) GL(M) admits a reduction to a bundle of Lorentz-orthonormal frames.
5) T(M) = L(M)⊕Σ(M) for some line bundle L(M) and some codimension-one vector bundle Σ(M).

By elementary obstruction theory [38, 49], when M is non-compact there is no obstruction to such a global section, but when M is compact the Euler-Poincaré characteristic $\chi(M)$ must vanish.

The choice of section L(M) is not canonical; indeed, it is equivalent to a choice of "rest frame." To be precise, we define the *rest frame* at $x\in M$ (mod L(M)) to be the *equivalence class* of Lorentz-orthonormal 4-frames in $T_x(M)$ that have one member that is a section of L(M). The local action of the Lorentz group on T(M), or GL(M) ([14]), or $\mathbb{R}T(M)$, for that matter, corresponds to making a different choices for L(M). Physically, one would expect these transformations to produce equivalent rest frames. One then necessarily wonders about how the space of sections of $\mathbb{R}T(M)$ partition into equivalence classes under this action and how they relate to the homotopy classes of such sections. This question goes beyond the scope of the present analysis and will be dealt with in a separate work.

---

[13] The quotation marks are supposed to remind one that this same sort of situation might involve PDE's for static-type problems, as well as dynamic ones.

[14] In terms of GL(M), we are talking about vertical isometric automorphisms.

Condition 5) suggests that we look into the issue of integrability [**44, 57, 59**]. As mentioned before, since L(M) is one-dimensional, it will be integrable; hence, there will be a one-dimensional foliation of M by the integral curves of L(M). (Often one demands that the these curves also be geodesics of the metric or the integral curves of a timelike Killing vector field.)

The question of whether $\Sigma(M)$ is integrable into codimension-one simultaneity leaves is more topologically involved. It is simpler to deal with this issue in a Lorentz manifold that has been given a "time orientation," which we now describe.

Now that we have a line bundle L(M) the next question to address is whether this bundle is orientable. This is equivalent to the existence of a global non-zero section of L(M), which amounts to a global non-zero time-like vector field. If such a vector field exists then M is said to be *time-orientable*. Notice that although the obstruction to the existence of *some* non-zero vector field is the same as the obstruction to the existence of the Lorentz structure, that does not imply that the vector field so obtained will be a section of L(M), or even homotopic to one.

The singularities that one might expect on a non-time-orientable Lorentz manifold would amount to the fixed points of the flow of a singular vector field, such as sources, sinks, vortices, as well as causal singularities of cosmological origin, such as closed timelike curves.

When a Lorentz manifold M is time-orientable and time-oriented by a vectorfield **t** then by means of *g* one can define the dual 1-form $\theta = i_v g$. This 1-form will have $\Sigma(M)$ as its annihilating sub-bundle. (Of course, we could have defined this sub-bundle without the time orientation.)

The Frobenius form for $\Sigma(M)$ is, of course, $\Phi = \theta \wedge d\theta$, and $\Sigma(M)$ will be integrable iff $\Phi = 0$. This is equivalent to saying that there is a 1-form $\eta$ such that $d\theta = \eta \wedge \theta$. A sufficient, but not necessary, condition for integrability is that $\theta$ be closed, $d\theta = 0$, or, in hydrodynamical terminology, *irrotational*. A sufficient, but not necessary, condition for this to be true is that there is a smooth function $\tau \in C^\infty(M)$ such that $\theta = d\tau$, i.e., that $\theta$ be exact; $\tau$ is sometimes called a "proper time function" for M.

To summarize, we have the three successively stronger conditions on $\theta$:
1. Integrability:   $d\theta = \eta \wedge \theta$,   for some $\tau \in \Lambda^1(M)$,
2. Closedness:   $d\theta = 0$,
3. Exactness:   $\theta = d\tau$,   for some $\tau \in \Lambda^0(M)$.

Only the strongest condition expresses the leaves of a codimension-one foliation in the form of level surfaces for a smooth function. In any event, the physical interpretation of the leaves are that they define a notion of simultaneity of events, as viewed in the rest frame that is defined by our choice of L(M), to begin with. This suggests that a "gauge transformation" that is associated with a change in the choice of **t** must correspond to a diffeomorphism M $\to$ M that is an isomorphism of the resulting codimension-one foliations, i.e., a diffeomorphism that preserves leaves. We shall not pursue this topic further here, as it goes beyond the scope of the present analysis.

Often causality considerations, such as the existence of maximal Cauchy hypersurfaces [**19, 28, 56**], demand that M must be trivially foliated as a product manifold $\mathbb{R} \times \Sigma$. In particular, this would make all of the leaves of the proper-time foliation diffeomorphic. The simplest way to introduce non-diffeomorphic leaves would be to weaken the assumption that there be a *maximal* Cauchy hypersurface, which has the effect of saying that the spacetime manifold is generated by the time evolution of a single proper time-like leaf. This might be as stringent a requirement on the topology of spacetime as demanding that it admit a single coordinate system; non-maximal Cauchy hypersurfaces would then be analogous to the local charts of the manifold atlas.

Ignoring the foliation by integral curves of **t** for the moment, we mention an aspect of proper-time evolution that has been discussed at length: the notion of *Lorentz cobordism* [**47, 59, 60**]*,* which is a convenient way of modeling "topology-changing processes" at the fundamental – i.e., cosmological – level. Rather than discuss the concept in detail, for the moment, we will simply show how it relates to the topic at hand of foliated cobordism.

By definition, two closed *n*-manifolds $S_1$ and $S_2$ are *Lorentz corbordant* – or *line field corbordant* – iff their disjoint union bounds an *n*+1-dimensional manifold V which carries a global line field L(V) that intersects the boundary components transversally. Since the line field L(V) can also be considered to define a one-dimensional foliation on V by way of integration, we see that any Lorentz cobordism is associated with the elementary type of foliated cobordism that was discussed above.

In the case of the proper time foliation of spacetime, $S_1$ and $S_2$ are simultaneity hypersurfaces $\tau_i$ and $\tau_f$ and the line field is spanned by the gradient vector field $\nabla \tau$. From Morse theory, if there is to be a change in topology between these two proper time instants, then $\nabla \tau$ must have a zero somewhere "between" them. This situation is represented in Fig. 2.

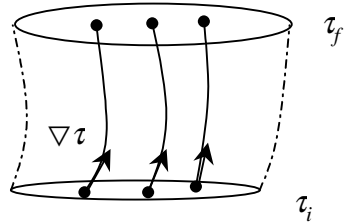

Fig. 2. Lorentz cobordism: non-singular and singular cases.

G. *Flow tubes and vortex tubes.* In continuum mechanics (mostly in hydrodynamics) [**3, 5, 32**], one encounters two notions that define elementary forms of foliated cobordism: *flow tubes*, which are called *world tubes* in relativistic hydrodynamics, and *vortex tubes.*

Actually, there are two distinct types of each, depending on whether one starts with a loop that does or does not bound a surface, i.e., a loop that is or is not contractible. If a loop $\Gamma$ does not bound, which is, of course, possible only if space is not simply connected, then if the integral curves of the flow of some velocity vector field **v** intersect

Γ transversally, one can use Γ and its evolutes up to any later proper time Γ($\tau$) to define a two-dimensional manifold T, with boundary $\partial T = -\Gamma \cup \Gamma(\tau)$, i.e., a foliated cobordism between Γ and Γ($\tau$). This defines one type of flow tube. Since T is foliated by streamlines, one says that T is a *streamlined surface.*

If Γ bounds a surface S that is also transverse to the flow then the interior of S and its evolutes up to any later proper time S($\tau$) define a foliated cobordism of one higher dimension, along with the foliated cobordism of the boundary as before. However, the space swept out by S with its boundary will not generally be a *differentiable* manifold, but only piecewise so, due to the way the lateral (timelike) boundary component intersects the spacelike component. This does not actually inhibit one from defining boundary integrals, but one must be careful in applying Stokes's theorem, since a boundary component might still have a non-vacuous "boundary" (although not in the strict homological sense).

The vorticity vector field ω that is defined by **v** also has a foliation by flow curves. If *u* is the co-velocity 1-form that is metric-dual to **v** then this vector field is defined by the dual of either the vorticity 2-form *du* in the non-relativistic case, or by *u*^*du* in the relativistic case:

$$i_\omega \eta = \begin{cases} du & \text{non - relativistic motion} \\ u \wedge du & \text{relativistic motion} \end{cases}.$$

Here, $\eta$ represents either the space or spacetime volume element, respectively, so we are implicitly assuming that the manifold in question is orientable. In either case, vorticity represents an obstruction to the integrability of the flow defined by *u* into a gradient flow.

The same considerations described in the context of flow tubes apply to the definition of analogous tubes, called *vortex tubes,* in terms of the flow of ω as well.

As we shall see later, Kelvin's and Helmholtz's theorem for flow tubes and vortex tubes define integral invariants that take the form of foliated cobordism invariants.

**IV. WAVE MOTION.** In physics, one must always be careful to distinguish between the theory of *physical waves* and the theory of the (linear) *wave equation*. Nowadays, with all of the attention being paid to nonlinear wave equations, this is easier to concede, but the possibility of confusion still exists. The basic objective of this article is to generalize the theory of physical wave mechanics beyond the particular wave equations by means of the methods of foliated cobordism. First, we examine the usual picture discussed in the theory of PDE's and then look for a way out.

A. *Cauchy problem for hyperbolic second-order PDE's.* There are various ways of representing second-order partial differential equations on manifolds. The path we shall take here is that of defining them by the level sets of differentiable functions on the space of 2-jets of $C^2$ functions on our manifold M:

$$F: J^2(M, \mathbb{R}) \to \mathbb{R}, \qquad (x^i, u, p_i, q_{ij}) \mapsto F(x^i, u, p_i, q_{ij}).$$

A solution of such an equation is a $C^2$ function *u* on M, whose 2-jet prolongation:

$$j^2 u = (x^i, u(x^i), \frac{\partial u}{\partial x^i}(x^i), \frac{\partial^2 u}{\partial x^i \partial x^j}(x^i)),$$

lies in the level set in question. Hence, since a 2-jet prolongation of a $C^2$ function $u$ satisfies:

$$u = u(x^i)$$
$$p_i = \frac{\partial u}{\partial x^i}$$
$$q_{ij} = \frac{\partial p_i}{\partial x^j} = \frac{\partial^2 u}{\partial x^i \partial x^j} = \frac{\partial p_j}{\partial x^i}$$

we must extend our previous canonical one-form on $J^1(M, \mathbb{R})$ to a set of $2n+1$ one-forms on $J^2(M, \mathbb{R})$:

$$\theta = du - p_i dx^i$$
$$\theta^i = dp_i - q_{ij} dx^j$$
$$\phi^i = (q_{ij} - q_{ji}) dx^j.$$

The exterior differential system defined by the equation $F = 0$ is then spanned by set of generators $\{F, \theta, \theta^i, \phi^i\}$ whose closure is then spanned by the set, $\{F, dF, \theta, d\theta, \theta^i, d\theta^i, \phi^i, d\phi^i\}$. This time, we have:

$$dF = F_{x^i} dx^i + F_u du + F_{p_i} dp_i + F_{q_{ij}} dq_{ij}$$
$$d\theta = dx^i \wedge dp_i$$
$$d\theta^i = dx^j \wedge dq_{ij}$$
$$d\phi^i = (dq_{ij} - dq_{ji}) \wedge dx^j.$$

A vector field on $J^2(M, \mathbb{R})$:

$$X = X^i \frac{\partial}{\partial x^i} + X_u \frac{\partial}{\partial u} + X_{p_i} \frac{\partial}{\partial p_i} + X_{q_{ij}} \frac{\partial}{\partial q_{ij}}$$

will be a *characteristic vector field* for the ideal generated by $\{F, dF, \theta, d\theta, \theta^i, d\theta^i, \phi^i, d\phi^i\}$ if it satisfies the following equations:

$$i_X dF = F_{x^i} X_i + F_u X_u + F_{p_i} X_{p_i} + F_{q_{ij}} X_{q_{ij}} = 0$$
$$i_X \theta = X_u - p_i X_i = 0 \qquad \text{so} \qquad X_u = p_i X^i$$
$$i_X d\theta = X_{p_i} dx^i - X_i dp_i \in I$$
$$i_X \theta^i = X_{p_i} - q_{ij} X_j = 0 \qquad \text{so} \qquad X_{p_i} = q_{ij} X_j$$
$$i_X d\theta^i = X^j dq_{ij} - X_{q_{ij}} dx^j \in I$$
$$i_X \phi^i = (q_{ij} - q_{ji}) X_j = 0 \qquad \text{so} \qquad q_{ij} X_j = q_{ji} X_j$$
$$i_X d\phi^i = (X_{q_{ij}} - X_{q_{ji}}) dx^j - X_j (dq_{ij} - dq_{ji}) \in I.$$

The condition that a 1-form be an element of I is that it be expressible in the form:
$\alpha dF + \beta \theta + \gamma_i \theta^i + \delta_i \phi^i =$
$[\alpha F_j - \beta p_j - \gamma_i q_{ij} + \delta_i (q_{ij} - q_{ji})] dx^j + [\alpha F_u + \beta] du + [\alpha F_{p_j} + \gamma_j] dp_j + \alpha F_{q_{ij}} dq_{ij},$
for some appropriate functions, $\alpha, \beta, \gamma_i, \delta_i$.

When we do the algebra for the system above, we get the equations:

$$\begin{aligned}
X_i &= \gamma_i \\
X_u &= p_i \gamma_i \\
X_{p_i} &= q_{ij} \gamma_j \\
X_{q_{ij}} &= -\alpha_i [F_j + F_u p_j + F_{p_k} q_{kj}]
\end{aligned}$$

and:

$$\gamma_j \, dq_{ij} = \alpha_i F_{q_{jk}} \, dq_{jk}$$
$$\gamma_j \, dq_{ji} = \bar{\alpha}_i F_{q_{jk}} \, dq_{jk}.$$

If we multiply each of the last two equations by $\gamma_i$ and sum then we should get the same thing as when we do that with the second equation:

$$\gamma_i \gamma_j dq_{ij} = (\alpha_k \gamma_k) F_{q_{ij}} \, dq_{ij} = (\bar{\alpha}_k \gamma_k) F_{q_{ij}} \, dq_{ij},$$

hence:

$$(\alpha_k \gamma_k) F_{q_{ij}} = \gamma_i \gamma_j,$$

which also gives:

$$(\alpha_k \gamma_k) F_{q_{ij}} \gamma_i \gamma_j = \gamma^4.$$

Since the vector field with components $X_i = \gamma_i$ must span a line in each tangent space, it cannot be zero anywhere. Hence, there are two conditions under which no solution $\alpha_i$ can be found:

$$\alpha_k \gamma_k = 0$$
$$\alpha_k \gamma_k \neq 0, \quad \text{but} \quad F_{q_{ij}} \gamma_i \gamma_j = 0.$$

The last condition takes the form of the vanishing of a quadratic form defined each tangent space $T_xM$ by $F_{q_{ij}}$ for some vector. The character of the PDE defined by $F = 0$ will then be determined by the character of its signature matrix $\text{diag}[\sigma_1, \ldots, \sigma_n]$ where $\sigma_k = -1, 0, +1$.

One defines the PDE in question to be *elliptic* iff all of the $\sigma_k = 1$. It is called *hyperbolic* iff $\sigma_1 = 1$ and the other $\sigma_k = -1$; this has the effect of singling out a preferred direction of "time evolution" for the solutions. If any of the eigenvalues are zero, one calls the PDE *parabolic*. Note that in the elliptic case, the only solution to the *characteristic equation* for the PDE:

$$F_{q_{ij}} \gamma_i \gamma_j = 0,$$

is zero, so one cannot use the method of characteristics to solve elliptic equations. In fact, one generally cannot pose the Cauchy problem for elliptic boundary-value problems, since it will generally be overdetermined, compared to the Dirichlet or Neumann problems.

In the hyperbolic case, a solution to the characteristic equation will not be unique, since we have defined the equation for an $n-1$-dimensional *characteristic cone* in each tangent space to M, which, in the case of spacetime is referred to as a *light cone* at that point. When we look at the $n-1$-dimensional submanifold of $J^2(M, \mathbb{R})$ that all of the characteristic cones collectively define we see we are dealing with an integral manifold to the rank-$n-1$ sub-bundle of $T(J^2(M, \mathbb{R}))$ that is spanned by all possible characteristic vector fields; this submanifold is called the *characteristic submanifold* of F.

The Cauchy problem for a second-order PDE must therefore be modified since we are now foliating a hypersurface in $J^2(M, \mathbb{R})$ by leaves of dimension higher than one. The Cauchy data for an $n-1$-dimensional initial submanifold $\Sigma$ of M must now include the normal derivative $u_n$ of $u$ in the Cauchy data, as well as the function $u|_\Sigma$. Furthermore, there are the two aforementioned possibilities for $u_n$: it can be characteristic or non-characteristic. In the first case, one cannot find the normal value of the second derivative of the solution from the Cauchy data, and the Cauchy problem will not generally be solvable. In the non-characteristic case, for which the initial normal derivative is not light-like, one can solve for the aforementioned normal second derivative and thus propagate the Cauchy data.

Although the characteristic system of our hyperbolic second-order PDE is not one-dimensional, as it was for the case of a first-order PDE, one can still speak of integrating a particular choice of characteristic vector field, i.e., one defined by a particular choice of $\gamma_i$. The resulting integral curves are called the *bicharacteristics* of the PDE, or the exterior differential system it defines. In the characteristic case, they will be essentially the "light rays" of the hyperbolic system, and will not be transverse to the characteristic manifold, but will lie in it. Of course, this is reminiscent of the way the characteristic curves of the Hamilton-Jacobi equation had to lie in the same submanifold of T*(M) that was defined by the appropriate level surface of the Hamiltonian. We shall discuss Hamiltonian optics after a few more remarks.

In the non-characteristic case, one is free to choose $\gamma_i$ and $\alpha_i$, except insofar as $\gamma_i$ cannot be light-like and $\alpha_i$ cannot be Euclidean-orthogonal to it.

For the non-characteristic case, if $\hat{\mathbf{n}}$ is the unit normal vector field to the initial submanifold $\Sigma$ then we envision a situation like Figure 3. When the normal derivative of our initial wave function is characteristic, i.e., light-like, we get Fig. 4.

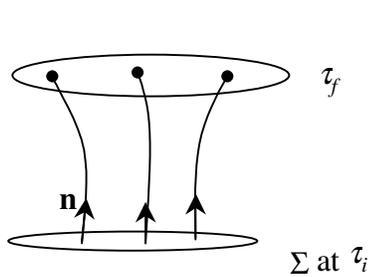 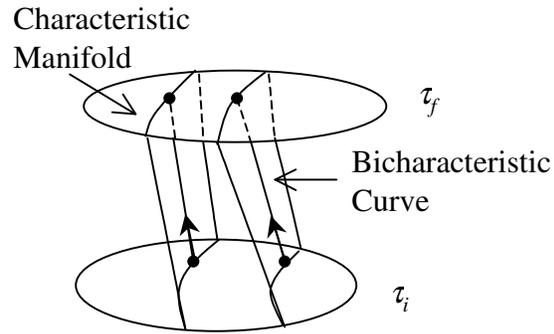

Fig. 3. Non-characteristic Cauchy data.    Fig. 4. Characteristic Cauchy data.

It is interesting that although the Cauchy problem is well posed only in the non-characteristic case, it seems that, nevertheless, physics and applied mathematics have had little to say about the specific nature of the non-characteristic solutions, beyond their existence. Perhaps this is because one generally expects that all physical waves will propagate with the same speed in a given medium, regardless of the circumstances of their generation; the main difference the initial speed will make is in the formation of shock discontinuities if that initial speed exceeds the speed of the medium. Even still, the discontinuities propagate along bicharacteristics [**26**].

B. *Optics.* One of the richest sources of inspiration for the analysis of wave equations is the study of optics [**6, 25**]. There are two main approaches to optics: wave optics and geometrical optics. Of course, wave optics is perhaps the intrinsic one, since it treats the motion of an electromagnetic wave in terms of boundary-value problems in the wave equations for the **E** and **B** fields that result from the Maxwell equations. Moreover, when one considers nonlinear optics, for which the wave equation is no longer the linear one, this is essential. However, the wave approach to optics is also generally quite complicated, at least for practical purposes – or at least needlessly tedious. In particular, it is amusing to see how something as elementary as the Law of Reflection can be deduced from the solution to an appropriate boundary-value problem in the wave equation.

Geometrical optics, which historically predated wave optics, is more straightforward for practical purposes, but also involves an approximation: one regards the wave fronts as infinitesimally close, so that the normals to those surfaces define a vector field that generates the rays as integral curves. This approximation amounts to either a high frequency or zero wavelength limit; it breaks down in precisely the quantum domain.

Not surprisingly, if one considers this latter image of wave hypersurfaces following orthogonal trajectories then the Hamilton-Jacobi approach is the definitive one, at least mathematically. Variationally, it starts with Fermat's principle: light rays follow the path of least time in the space manifold $\Sigma$. This means they extremize the action functional defined by the *optical length* of the path:

$$S[\gamma] = \int_\gamma n(\tau, x^i(s), \dot{x}^i(s)) ds = \int_\gamma n(\tau, x^i(\tau), \dot{x}^i(\tau)) \sqrt{g_{ij} \dot{x}^i \dot{x}^j}\, d\tau.$$

The Lagrangian density $\mathcal{L} \in C^\infty(J^1(\mathbb{R}, \Sigma))$ takes the form:

$$\mathcal{L}(\tau, x^i, \dot{x}^i) = n(\tau, x^i, \dot{x}^i) \sqrt{g_{jk} \dot{x}^j \dot{x}^k},$$

in which the *medium function* $n$ equals the index of refraction $\frac{c_0}{c}$ of the medium. In effect, the variational problem we have defined is conformal to the geodesic problem of the space metric, by way of the conformal factor $n$. Hence, the extremals will be geodesics of the conformally transformed metric $n^2 g$. A similar situation is defined by the relativistic dynamics of holonomic fluids [**34**], for which the only differences are the extra time dimension in the velocity, the four-dimensional metric, and the interpretation of $n$ as a hydrodynamical property of the medium instead of an optical one.

Of course, to be fully relativistic about light rays, we need to add the time dimension $t$. If we assume that it is related to the three-dimensional arc-length by:

$$ds = c\, dt,$$

then we get:

$$d\tau^2 = c^2 dt^2 - ds^2 = 0.$$

This makes the space geodesics into *null geodesics* in spacetime. It also implies that one cannot use proper time to parameterize these null geodesics. Hence, one generally introduces a different parameterization, such as the "affine parameter."

In order to put the foregoing into Hamilton-Jacobi form, we first form the Poincaré-Cartan 1-form:

$$\theta = p_i dx^i - H d\tau,$$

and assume that $n = n(x^i(s))$ is isotropic and independent of time. This time, we have used the Lagrangian that is closely related to the optical length:

$$\mathcal{L} = \tfrac{1}{2} n^2 g(\mathbf{v}, \mathbf{v}) \equiv \tfrac{1}{2} \bar{g}(\mathbf{v}, \mathbf{v}),$$

which defines the *energy* of the curve and produces the Hamiltonian:

$$H = \frac{\partial \mathcal{L}}{\partial v^i} v^i - \mathcal{L} = n^2 g_{ij} v^j v^i - \tfrac{1}{2} n^2 g_{ij} v^j v^i = \tfrac{1}{2} n^2 g(u, u) \equiv \tfrac{1}{2} \bar{g}(u,u).$$

The Hamilton-Jacobi equation that follows for $E = 1$:

$$\bar{g}(dS, dS) = 1, \quad \text{or} \quad g(dS, dS) = n^{-2},$$

is called the *eikonal equation*. The level surfaces of its Hamilton characteristic function S are called *wave fronts*. The bicharacteristic vector field on $T^*\Sigma$ is then the Hamiltonian vector field for H:

$$X_H = \frac{\partial H}{\partial u_i} \frac{\partial}{\partial x^i} - \frac{\partial H}{\partial x^i} \frac{\partial}{\partial u_i} = \bar{g}^{ij} u_i \frac{\partial}{\partial x^i} - \frac{1}{2} \frac{\partial \bar{g}}{\partial x^i}(u,u) \frac{\partial}{\partial u_i},$$

whose integral curves are solutions of the ODE's:

$$\frac{dx^i}{d\tau} = \bar{g}^{ij} u_i, \quad \frac{du_i}{d\tau} = -\frac{1}{2} \frac{\partial \bar{g}}{\partial x^i}(u,u).$$

The first equation is really an identity, and the second one is the Hamiltonian form of the geodesic equation for the conformally transformed metric $\bar{g}$. The projection of the bicharacteristic curves along a geodesic section $dS$ of $T^*\Sigma$ give the geodesic curves described before, which are usually called *light rays*.

The way that geometrical optics relates to foliated cobordism is then immediate once we have put the problem into Hamilton-Jacobi form.

C. *De Broglie waves* [**13, 51**]. A significant step in the early years of quantum mechanics was the suggestion of de Broglie that massive particles had wave-like properties, just as massless particles (in the form of photons, back then) did. The now-classic relations he defined said that a point-like particle of energy E and momentum $p$ is equivalent to a wave with the following frequency and wavelength ([15]):

$$v = \frac{E}{h}, \quad \lambda = \frac{h}{p}.$$

These relations can be given a concise, relativistically invariant form if we introduce the radial frequency $\omega = 2\pi v$ and wave number $\kappa = \frac{2\pi}{\lambda}$ and form the *frequency-wavenumber 1-form*:

$$k = \frac{\omega}{c} d\tau + \kappa \hat{n},$$

---

[15] Although the use of "natural" units, in which $c = \hbar = 1$, has a long and proud tradition in mathematical physics, it is in the study of wave mechanics, in which $c$ might depend on other things and $\hbar$ has an ambiguous significance that demands deeper physical interpretation, that the tradition becomes inappropriate.

in which $d\tau$ is the time-orientation 1-form and $\hat{n}$ is the unit 1-form that is spatially normal to the wave front. If $p = \dfrac{E}{c} d\tau + P\hat{n}$ is the energy-momentum 1-form for the particle then the de Broglie relations amount to the postulate that:
$$p = \hbar k.$$

Since this implies that:
$$p^2 = \eta(p, p) = (m_0 c)^2 = \hbar^2 k^2 = \text{Lorentz constant}$$
in which $m_0$ is the rest mass of the particle, we can identify a wave analogue to the rest mass in the form of the *Compton wave number*:
$$k_0 = \frac{m_0 c}{\hbar}.$$

Notice that since we have just observed that the energy-wavenumber 1-form for massive matter waves is timelike, not lightlike, we see that in the eyes of hyperbolic second-order PDE's, it should propagate as a non-characteristic solution.

Because of this "wave-particle duality," the early attempts at formulating a wave equation for massive particles were concentrated around trying to go beyond the "wave-ray duality" of Hamiltonian optics and Hamilton-Jacobi theory in such a way that one would retrieve geometrical mechanics from wave mechanics in the limit as $\hbar \to 0$, or, more precisely, the limit of vanishing wavelength (i.e., high frequency).

The relativistic wave equation that quantum mechanics settled upon was defined for a complex-valued wave-particle $\psi$ with no intrinsic angular momentum – i.e., spin – namely, the Klein-Gordon equation:
$$\Box \psi + k_0^2 \psi = 0,$$
in which $\Box = \eta^{ij} \dfrac{\partial}{\partial x^i} \dfrac{\partial}{\partial x^j}$ is the d'Alembertian operator ([16]).

The reason that far more physical problems have been worked through for its non-relativistic analogue – the Schrödinger equation – is that the physical interpretation that was given to wave mechanics was the statistical interpretation of Born and the Copenhagen School. The Klein-Gordon equation, although relativistically invariant, nevertheless defined a current that, due to its possible negative values, was not consistent with a probability current interpretation. Objectively, in the eyes of *reductio ad absurdum* this could also serve as a contradiction to the validity of the statistical interpretation, as well. However, in the eyes of the Correspondence Principle of physics, before you contradict something that has satisfied most of the physicists most of the time, you need a better theory to replace it with.

A significant feature of the Klein-Gordon equation is that if the only thing about $\psi$ that has any physical significance is its modulus-squared $\psi\psi^*$, which defines the probability density function for the location of the particle, then the phase angle that appears in $\psi$ appears to be, on the surface of things, physically irrelevant. If one allows the choice of zero phase angle at one point to be independent of the choice at another then one must introduce a U(1) gauge structure – i.e., a U(1) principle bundle and a connection on it – in order to make the system defined by the wave invariant under all such arbitrary assignments. Interestingly, electromagnetism, which describes massless waves, among other things, also has such a U(1) invariance.

---

[16] One should note, as O. Klein did [**51**], that if one extends to a five-dimensional d'Alembertian by adding another spacelike dimension then the Klein-Gordon equation takes the form of the equation that one obtains by separating the fifth variable out of the wave function, much as one obtains the Helmholtz equation $\Delta \psi + k^2 \psi = 0$ from the four-dimensional wave equation by separating out the time variable; there was some follow-up work done at the time into the subject of "5-optics."

Since traditional probability and statistics says little, if anything, about representing a probability density function by the modulus-squared of a complex function, one really ought to regard the introduction of this U(1) principle bundle as somewhat formal from a physical standpoint. It would be more intuitively satisfying to represent this bundle in terms of things that are intrinsically associated with the spacetime manifold itself, such as the bundle of linear frames GL(M). We shall now discuss how wave motion in general defines a reduction of GL(M) to an SO(2)-principle sub-bundle, i.e., an SO(2) structure.

D. *Generalization.* So far, we have discussed three basic pictures for wave motion: the general methods of hyperbolic second-order PDE's, methods of geometrical optics for massless waves, and the basic notions of wave mechanics for massive waves. Obviously, one would like to find a general picture that includes these cases as consequences.

The first conclusion that this article has been leading up to is that wave motion can defined by a foliated cobordism between spacelike closed submanifolds $\tau_i$, $\tau_f$ with foliations of codimension one in a Lorentz manifold. These two submanifolds will represent two proper time simultaneity leaves of spacetime, called *isochrones,* relative to a choice of rest-frame and the timelike connecting leaves will represent the *isophase* hypersurfaces for the wave. The leaves of the spacelike submanifolds are called *momentary fronts*. This situation is schematically represented in Fig. 5.

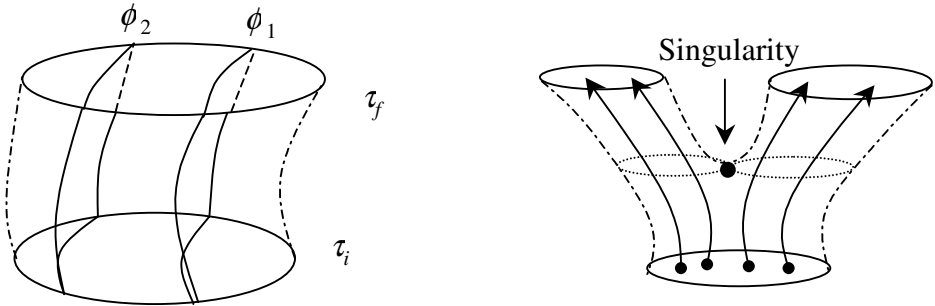

Fig. 5. Wave motion as foliated cobordism.

The second conclusion of the present analysis is that if the foliation of a simultaneity leaf by momentary fronts is defined a spacelike unit 1-form *n,* which represents the space direction of propagation of the momentary fronts, and the proper-time foliation is defined by the timelike unit 1-form $\theta$ that one obtains from the Lorentz structure and which is orthogonal to *n,* then the orthonormal 2-frame field $\{\theta, n\}$ defines a reduction of the bundle of linear frames GL(M) to an SO(2) principle bundle SO(2)(M), i.e., a G-*structure* on M with G = SO(2). Basically, the fiber of SO(2)(M) at any point $x \in M$ consists of all the orthonormal 4-frames at *x* that include $\theta$ and *n* as members. Since this leaves the other two members indeterminate, except for their plane of definition and length, this shows that SO(2) – or, equivalently, U(1) – gauge invariance is intrinsic to four-dimensional wave motion in general, at least as we have defined it.

In the case of electromagnetic waves, the plane spanned by this 2-frame contains the **E** and **B** vectors, and a common choice of orthonormal 2-frame consists of the real parts of the polarization vectors. However, one does not usually regard the U(1) gauge invariance of the electromagnetic field as being defined by the orientation of these vectors, but by the ambiguity in defining the electromagnetic potential 1-form.

Although the foregoing abstractions would probably be sufficient for a purely mathematical study, they are also too general to suggest any deep physical consequences, except by distant corollaries. Hence, we need to add a few more restrictions in order to move closer to a physically useful generality in the name of wave motion.

First, we address the unavoidable question of defining what we are calling a "wave" in terms of natural phenomena. This is not as mathematically trivial as it sounds when one starts to include modern considerations such as nonlinearity, the non-existence of translational invariance for a general Lorentz manifold, and the fact that physical waves do not have to be either periodic, infinite in extent, or lightlike, in general.

We start by saying that a *wave* on a Lorentz manifold M consists of a section *s* of a vector bundle E → M and a *wave structure* for *s*. A wave structure for *s* is then defined by a pair of non-zero 1-forms $\theta$ and $\phi$, such that:
1) The closure of their common domain of definition equals the support of *s*.
2) $\theta$ is a unit timelike 1-form that is metric-dual to a vector field **t** that generates the line field that defines the Lorentz structure.
3) $\phi$ is linearly independent of $\theta$ and is either lightlike everywhere on supp(*s*) or spacelike everywhere on supp(*s*).

In particular, the closure of the above domain is assumed to be compact. (Notice that this means M can still have topological obstructions to the existence of *global* 2-frames.)

Although there is a very strong temptation to broaden the domain of definition of a wave to all of M, especially when one considers cosmological problems, such as the initial-value formulation of Einstein's equations, one should resist that impulse, since topology is very strict in the differences between compact and non-compact spaces; in particular, this is true for Lorentz structures. However, the physical reality is that a wave is always produced by a finite amount of energy that acts for a finite length of time. The consequence of this is that the momentary fronts can only be bounded in their extent, both in space and in time. The fact that we are not demanding that the domain of definition of the wave structure for a given section be necessarily closed is simply a way of leaving open the possibility that the wave might have singularities, such as sources, that lie outside of its domain of definition as limit points.

Starting with the pair $\theta, \phi$, we then begin to examine the various geometrical consequences in terms of wave motion. The vector field **t** that is metric-dual to $\theta$ generates a one-dimensional foliation of timelike curves that we shall interpret as the proper time flow. The pseudo-norm of $\theta$:

$$\|\theta\| = \sqrt{|g(\theta,\theta)|},$$

represents the speed of the parameterization of these timelike curves; hence, our assumption that it is identically 1.

From $\theta$ and $\phi$, we construct two more 1-forms: the unit normal to the momentary fronts:
$$n = \text{normalize}(f - g(\phi, \theta)\theta),$$
and the *ray covelocity* 1-form:
$$u = -g(\phi, n)\theta + g(\phi, \theta)n.$$
Note that $u^2 = -\phi^2$. Hence, since $\phi$ is spacelike, $u$ is timelike.

Insofar as a choice of timelike line through the origin in a tangent space also defines a light cone in that tangent space, we can obtain the *characteristic speed* of the wave structure as the slope of any lightlike line through the origin; by Lorentz invariance, it will be the same for any lightlike line. If **l** is any lightlike vector then we can form the orthogonal expansion of **l** relative to **t**:
$$\mathbf{l} = g(\mathbf{l}, \mathbf{t})\mathbf{t} + \mathbf{l}_s,$$
and define the characteristic speed at $x$ to be:
$$c(x) = -\frac{g(\mathbf{l}_s, \mathbf{l}_s)}{g(\mathbf{l}, \mathbf{t})}.$$

Although it is customary to set $c = 1$ identically by a judicious choice of units, this is, of course, only valid for a wave structure in vacuo and eliminates all consideration of refraction. In effect, the expression for $c(x)$ will depend on the choice of spacelike Riemannian metric $g_s$ that one uses to construct the Lorentz metric from the chosen timelike line:
$$g = \mathbf{t} \otimes \mathbf{t} - g_s.$$
For the case of an isotropic medium, the index of refraction will take the form of a conformal factor on the spacelike metric.

If we regard the ratio of the spacelike component of $\phi$ at a point $x$ to its timelike component as $1/c$ times the *phase velocity* $u(x)$ of the momentary wave front at $x$:
$$\frac{u}{c} = \frac{g(\phi, n)}{g(\phi, \theta)},$$
then minus the corresponding ratio for the ray covelocity, which defines what we call the *ray velocity* $v(x)$, becomes:
$$\frac{v}{c} = \frac{g(\phi, \theta)}{g(\phi, n)}.$$
If we multiply the two expressions then we get the classic de Broglie relation:
$$uv = c^2.$$
Hence, the phase velocity and the ray velocity will be the same iff the isophase is characteristic.

The unit covector field $n$ defines two more useful objects: Its metric-dual vector field **n** generates a one-dimensional foliation in each simultaneity leaf that amounts to the spacelike *rays* of the wave structure. When combined with $\theta$, the pair $\{\theta, n\}$ defines a timelike orthonormal 2-frame field on the wave structure, hence, an SO(2)-structure on it.

It might seem straightforward to define a frequency-wavenumber 1-form once one has defined the 2-frame $\{\theta, n\}$. However, if one considers that physical waves can include moving shapes with no clear periodicity to them, such as square pulses in optical fibers, then the physical nature of both frequency and wave number become ill defined. The usual argument that one can always expand the more amorphous wave shapes into a Fourier series in terms of periodic functions falls through in a more general context since manifolds do not generally admit an action of the translation group, or, equivalently, a "radius vector field," so spatial periodicity is harder to define, and, furthermore, if the equation that the section *s* solves is nonlinear, the utility of such an expansion is limited ([17]).

First, we assume that the Levi-Civita connection on SO(3,1)(M) has also been extended to a connection on E then we can use the parallel translation along the integral curves of **t** and **n** to define what we would expect of $\omega$ and $\lambda$, in place of the vector space translations that we have surrendered by going to a more general manifold. In particular, the *period* T($x$) of *s* at $x \in M$ – if it exists – is defined to be the smallest positive real number such that $s(\gamma(\tau+T)) = s(\gamma(\tau))$, in which equality implies a parallel translation, unless *s* is a scalar function, and $x = \gamma(\tau)$, for some parameterized integral curve of **t**.

To define the corresponding *wavelength* $\lambda(x)$ of *s* at *x*, we use the spacelike vector field **n** to generate a flow in each simultaneity leaf. Along a curve $\alpha$ of this foliation, which we parameterize by arclength *l*, we then make a similar definition that $\lambda(x)$ is the smallest positive number that makes $s(\alpha(l + \lambda)) = s(\alpha(l))$, with the same clarifications as before.

Since both T($x$) and $\lambda(x)$ are non-zero, we can define the corresponding angular frequency and angular wave number at *x*:

$$\omega(x) = \frac{2\pi}{T(x)}, \qquad \kappa(x) = \frac{2\pi}{\lambda(x)},$$

and the 1-form at *x*:

$$k|_x = \frac{\omega(x)}{c(x)} \theta|_x + \kappa(x) n|_x.$$

It is tempting to simply define $c(x) = \lambda(x)/T(x) = \lambda(x)f(x)$, instead of our former definition, but this oversimplifies the issue of dispersion laws, since it only gives the linear law $\omega(x) = c(x)\kappa(x)$. If one expects to see a deformation of the shape of a wave envelope during its evolution, one must generally include dispersion; however, an invariant shape might still involve dispersion, as is the case with solitons, for which the effects of nonlinearity balances out the effects of dispersion.

In order to extend $k|_x$ to a 1-form *k* that is defined over the entire wave structure, we must first assume that $\omega$, $\lambda$, and *c* exist at every point of the support of *s*. We then say that the pair $\{\theta, \phi\}$ defines a *periodic wave structure* for *s*. Furthermore, we need to assume that both $\theta$ and *n* are exact – $\theta = d\tau$, $\phi = dS$ – in order to extend the values of $\omega$ and $\lambda$ transversally across the sections of their flows; this makes the distribution spanned by the 2-frame field $\{\theta, n\}$ integrable.

An important issue that one must now consider is whether a different choice of rest frame, i.e., Lorentz structure, will give a different 1-form for *k*. This is easily seen to not be the case, since a different choice of Lorentz structure will be related to the first one by a Lorentz transformation. This will, in turn, act on the orthonormal 2-frame $\{\theta, n\}$ to produce another orthonormal 2-frame. However, both of the components of *k* are metric-defined quantities that will transform covariantly to compensate for the change in frame.

---

[17] Considerable progress has been made along these lines under the purview of the theory of Fourier integral operators on manifolds (cf. [**52, 53**]).

In the case of a periodic wave structure, one can use the de Broglie relations to simply *define* the energy-momentum 1-form $p$ that corresponds to $k$ as $p = \hbar k$. In the aperiodic case, unless one is allowed some sort of "Fourieresque" expansion into periodic components, which sounds suspiciously linear in character, one must consider the construction of $k$ to be a more debatable issue.

Once one has such a $k$ or $p$, one is free to discuss the significance of their Lorentz pseudo-norms $k_0^2 = g(k, k)$ and $p_0^2 = g(p, p)$. We continue the tradition of identifying $k_0$ with the Compton wave number of the wave defined by $s$ and $p_0 = m_0 c$ with its rest mass density times $c$; similarly, $\hbar\omega$ is its energy density and $\hbar\kappa$ is the magnitude of its momentum density.

We conclude the discussion of our definition of a generalized wave structure with the observation that, except for $c$, the fundamental objects from which all of the physical definitions arise are the parameterizations $\tau$ and $l$, or, equivalently, the functions $\tau$ and $\phi$. A reparameterization of $\tau$ or $\phi$ defines a positive function on the wave structure, which, in turn, defines a conformal change of the Lorentz metric. In that light, the well known importance of the conformal group in wave theory – at least for the lightlike, i.e., massless, case – is seen to be related to questions of invariance under reparameterizations of the one-dimensional leaves of the foliations in question. In such a case, the distinction between a one-dimensional leaf of the foliation defined by a differential system and an integral curve of a vector field becomes crucial.

E. *Typical momentary wave-front foliations.* The most common type of momentary wave-front foliation of space Σ is by *plane waves*, which we understand to mean, in general, a codimension-one foliation whose leaves are all diffeomorphic to $\mathbb{R}^2$. This is actually a strong condition on a three-manifold, at least when it is orientable. We have the following theorem [**17**]:

**Theorem:**
> *Suppose that Σ is an orientable three-manifold (without boundary) that is foliated by leaves that are diffeomorphic to $\mathbb{R}^2$.*
> *i)    If Σ is non-compact and simply connected then it is diffeomorphic to $\mathbb{R}^3$;*
> *ii)   If Σ is compact then it is diffeomorphic to the three-torus $\mathrm{T}^3$.*

The canonical pathology of codimension-one foliations of compact three-manifolds with boundary is that of the *Reeb foliation* of the solid torus $\mathrm{D}^2 \times \mathrm{S}^1$, whose leaves will all be $\mathbb{R}^2$, except for the boundary. If one identifies the boundary points of two such foliated solid torii then one obtains a codimension-one foliation of $\mathrm{S}^3$ by planes.

If one wishes to loosen the restriction that the leaves be $\mathbb{R}^2$ to the restriction that they all be non-compact then we have:

**Sullivan's theorem:**
> *If Σ is closed and orientable and admits a codimension-one foliation by non-compact leaves then Σ admits a Riemannian metric for which the leaves are all minimal submanifolds.*

Such foliations are called *geometrically taut*. Their existence entails a restriction on the homotopy groups of $\Sigma$:

**Novikov's theorem:**
> *If $\Sigma$ is closed and orientable and admits a codimension-one foliation whose leaves are all non-compact then $\pi_2(\Sigma) = 0$.*

A partial converse to this is the theorem of Schoen, Yau, Sacks, and Uhlenbeck [**43, 46**] that if $\Sigma$ is a closed orientable Riemannian manifold with $\pi_2(\Sigma) \neq 0$ then it admits a minimal embedding of $S^2$.

It is interesting to note that plane waves − although commonplace mathematical devices to utilize − have no physical reality since their lack of compactness implies that they carry an infinite energy, and they must be generated by an infinite planar source. Generally, plane waves are only useful in the context of boundary-value problems that are posed in compact regions, such as parallelepipeds.

Another elementary wave foliation is by *spherical waves*, which we interpret to mean leaves that are diffeomorphic to $S^2$; of course, this also includes such deformations as elliptical waves. For such foliations we have:

**Reeb Stability Theorem:**
> *If a closed three-manifold $\Sigma$ admits a codimension-one foliation with at least one leaf that is diffeomorphic to $S^2$ then all of the leaves are either $S^2$ or $\mathbb{R}P^2$, and $\Sigma$ is either $S^2 \times S^1$ or the connected sum $\mathbb{R}P^3 \# \mathbb{R}P^3$.*

(The connected sum of two manifolds is formed by removing an open ball from each and identifying the boundaries thus created.) Note that spherical waves are generally assumed to emanate from a point source, which represents a singularity of such a foliation.

The case of cylindrical waves involves an interesting aspect in the form of whether one expects the line source to be open or closed. In the former case, the leaves would be diffeomorphic to $S^1 \times \mathbb{R}$; in the latter, they would be $T^2$.

One reason why Thurston's result on the continuous infinitude of inequivalent (under foliated cobordism) codimension-one foliations of $S^3$ might be of particular interest to wave motion is that one might regard the ischrones as 3-spheres as a first approximation to spatial topology. Physically, this would correspond to assuming that any wave will vanish at infinity on $\mathbb{R}^3$, which is essentially a finite-energy assumption on the wave. What Thurston said about asymptotically vanishing waves in $\mathbb{R}^3$ then is that there is a continuous infinitude of inequivalent ways in which space can be foliated into momentary fronts.

V. **FOLIATED COBORDISM INVARIANTS.** The invariants that we will discuss take the form of either differential or integral invariants, i.e., differential forms that are defined on each of a pair of foliated cobordant closed manifolds or their integrals over those manifolds. Because of the topological nature of foliated cobordism, closed differential invariants, which define de Rham cohomology classes, play an important role.

A. *Extended particle motion.* In the case of extended particle motion, for which foliated cobordism is essentially Lorentz or line cobordism, an obvious place to look for differential invariants is to look for differential forms that are constant along the flow that connects the boundary components. If $\alpha$ is such a form and **v** is a vector field that generates the flow then the latter condition can be expressed in terms of the Lie derivative:

$$L_\mathbf{v}\alpha = di_\mathbf{v}\alpha + i_\mathbf{v}d\alpha = 0.$$

If $\alpha$ is a closed $k$-form then $L_\mathbf{v}\alpha = d(i_\mathbf{v}\alpha)$ is exact. Hence, the de Rham cohomology class of $\alpha$, $[\alpha]$, is invariant under the flow. Of course, this is simply the statement that $H^*(\Sigma, \mathbb{R})$ is invariant under diffeomorphisms, which is what the flow of **v** consists of.

If we consider, more generally, a Lorentz cobordism whose flow is singular, the cobordism might no longer produce a diffeomorphism of the boundary components and the closed forms might no longer be invariant. However, as pointed out before, if the closed forms are of the same dimension as the boundary components then their integrals over the cobordant boundary components are still invariants, regardless of the flow.

Following the terminology of Cartan [9] − by way of Godbillon [20] − we call a $k$-form $\alpha$ an *absolute differential invariant* of the differential system spanned by **v** iff $L_{f\mathbf{v}}\alpha = 0$ for all smooth functions $f$. Note that this is a stronger condition than merely requiring the vanishing of $L_\mathbf{v}\alpha$ since we are requiring that this fact be independent of the parameterization of the integral curves. The definition means:

$$0 = di_{f\mathbf{v}}\alpha + i_{f\mathbf{v}}d\alpha = df \wedge i_\mathbf{v}\alpha + f(di_\mathbf{v}\alpha + i_\mathbf{v}d\alpha) \quad \text{for all } f \in C^\infty(M).$$

Since $f$ is arbitrary, this implies that we must have $i_\mathbf{v}\alpha = 0$, which then implies that $i_\mathbf{v}d\alpha = 0$. Hence, $\alpha$ is an absolute differential invariant of the flow of **v** iff:

$$i_\mathbf{v}\alpha = 0 \quad \text{and} \quad i_\mathbf{v}d\alpha = 0.$$

A weaker condition on $\alpha$ is to require that $d\alpha$ be an absolute differential invariant, in which case, one calls $\alpha$ a *relative differential invariant.* This is then equivalent to the single condition:

$$i_\mathbf{v}d\alpha = 0.$$

A natural consequence of the nature of Hamiltonian flows on symplectic manifolds is that the energy density of the system, as represented by H, will automatically be conserved, since:

$$L_{X_H} H = i_{X_H} dH = dH(X_H) = \Omega(X_H, X_H) = 0.$$

This is related to the fact that $\theta$ is a relative differential invariant of $X_H$, i.e., the symplectic form $\Omega = d\theta$ is an absolute differential invariant. This is equivalent to the statement that the flow of a Hamiltonian vector field preserves the symplectic form $\Omega$.

Suppose that M is orientable and time-orientable, with a volume element $\eta$ and a timelike unit vector field **t**. From this one can define a spatial volume element on each simultaneity leaf by:

$$\eta_s = i_\mathbf{t}\eta.$$

By definition, the flow of **t** is *spatially* incompressible iff $L_\mathbf{t}\eta_s = 0$, and *incompressible* iff $L_\mathbf{t}\eta = 0$. Since $i_\mathbf{t}\eta_s = 0$, this says that $\eta_s$ is an absolute differential invariant of **t** iff $i_\mathbf{t}d\eta_s = 0$. If our spatial volume element takes the form $\eta_s = dx\wedge dy\wedge dz$ for some coordinate system, this is assured, but if we have, more generally, $\eta_s = \rho\, dx\wedge dy\wedge dz$ then this would depend on the nature of $\rho$. In other words, conservation of volume does not have to imply conservation of mass.

Generally, the momentum 1-form *u* must be a *relative* invariant of a Hamiltonian flow, since it is the pull-down of the canonical 1-form $\theta$ on T*(M) by a section, and $\theta$ is a relative invariant of such a flow. Hence, it is the *dynamical vorticity* 2-form *dp* that must be an absolute invariant. In order to make *p* an absolute invariant one must assume that there is translational symmetry to the system that is described by H. We shall say more about that shortly.

Closely related to the differential invariants of a flow are its *integral invariants*. The difference is that if $S_1$ and $S_2$ are any two disjoint closed orientable *k*-submanifold that intersects the flow of **v** transversally then the *k*-form $\alpha$ is an *absolute integral invariant* of the flow of **v** iff:

$$\int_{S_1} \alpha = \int_{S_2} \alpha.$$

$\alpha$ is a *relative integral invariant* of the flow of **v** iff $d\alpha$ is an absolute integral invariant. An absolute integral invariant is then an elementary type of foliated cobordism invariant.

By the "forgetful functor," since foliated cobordism implies some more general forms of cobordism, namely, unoriented and oriented cobordism, the topological invariants of those forms of cobordism will have to apply, along with the ones more specific to the foliations. Some of them can be represented as integrals of closed differential forms. In particular, the Euler number for a compact oriented manifold can be represented as the integral of the Euler class of that manifold and if the manifold is of dimension 4*k* then its Pontrjagin numbers can be represented as integrals of various polynomials in the Pontrjagin classes of that manifold.

The earliest and most definitive works on integral invariants were by Poincaré [**39**] and Cartan [**9**]. Poincaré showed that the 1-form $p_i dx^i$ is a relative invariant of a Hamiltonian flow for the case of a time-invariant Hamiltonian when $p_i = \dfrac{\partial \mathscr{L}}{\partial \dot{x}^i}$, and Cartan extended that result to the 1-form $\theta = p_i dx^i - H d\tau$ for the time-varying case.

In fact, Cartan went further than the mere identification of an integral invariant; he showed that Hamilton's Principle of Least Action was *equivalent* to the postulate that the motion of a natural system is along trajectories that render the Poincaré-Cartan 1-form invariant. To obtain the equations of those trajectories, one simply looks for the characteristic system of $\theta$. However, throughout the ages this aspect of Cartan's theory has been lost, in favor of the more popular modern notion that Cartan ever said that Hamiltonian mechanics is best formulated on symplectic manifolds. Mathematically, one still reaches the same equations of motion in either case, but philosophically what Cartan was proposing was deeper than mere mathematically convenience.

If *u* is the covelocity 1-form corresponding to a velocity vector field **v** and $C_1$ and $C_2$ are disjoint loops that intersect the flow of **v** transversally – hence, they define a foliated cobordism – then the theorem of classical hydrodynamics called Kelvin's Theorem amounts to the statement that circulation is an absolute integral invariant, or, for that matter a foliated cobordism invariant. When the vector field is ω, the vorticity vector field dual to *du*, the same sort of situation is called Helmholtz's theorem. This amounts to the statement that circulation is also an absolute integral invariant of the flow of ω.

The situation in which the circulation of a closed covelocity 1-form – hence, an irrotational flow – is non-vanishing deserves special consideration. The only way this can happen is if the loop C around which one integrates does not bound a surface, so Stokes's theorem does not apply. This, in turn, implies that the manifold, M, cannot be simply connected. If the obstruction to the contraction of C to a point is a curve then such a flow is said to define a *line vortex*. The flow is also characterized by a covelocity that is closed, but not exact; hence, there would be no stream function.

One might also consider flow invariants that are not generally represented as differential forms ([18]), such as the metric tensor *g*. If the vector field **v** has the property that $L_\mathbf{v} g = 0$, then **v** is called a *Killing vector field*. This implies that the flow of **v** is by isometries, which amounts to a sort of rigid motion. A weaker condition, that $L_\mathbf{v} g = \lambda g$, for some smooth function $\lambda$ makes **v** a *conformal Killing vector field* ([19]). More to the point, one might consider the restriction $g_s$ of *g* to a spacelike three-dimensional closed submanifold of M. Since one can then express *g* in the form:

$$g = \theta \otimes \theta - g_s,$$

where $\theta$ is a timelike 1-form such that $\theta(\mathbf{v}) = -1$, one sees that in order for $g_s$ to be an invariant of the flow of **v**, one must have:

$$L_\mathbf{v} g = L_\mathbf{v}\theta \otimes \theta + \theta \otimes L_\mathbf{v}\theta. = i_\mathbf{v}d\theta \otimes \theta + \theta \otimes i_\mathbf{v}d\theta.$$

If **v** is irrotational its flow preserves the spatial metric iff it preserves the spacetime metric.

The classical result of the calculus of variations that goes by the name of Nöther's theorem can be given a formulation in terms of foliations [**48**].

First, one weakens the definition of a symplectic manifold to the definition of a *presymplectic manifold* M for which the closed 2-form Ω is no longer necessarily non-degenerate, but simply has constant rank. Then one observes that the sub-bundle of T(M) that is defined by the annihilating subspaces of Ω is integrable since Ω is closed. One calls this foliation the *characteristic foliation* defined by Ω ([20]).

Let G be a Lie group. The dual 𝔊* of its Lie algebra 𝔊 can be regarded as either linear functionals on 𝔊 or left-invariant 1-forms on G, which are sometimes called *Maurer-Cartan forms*. In geometrical mechanics, its elements are sometimes referred to as *torsors,* since they can represent forces or torques, depending on the physical nature of G.

Suppose that G acts canonically on the (pre)symplectic manifold M; hence, the form Ω is preserved by the action of G. A *momentum map* for the action of G on M is a map:

---

[18] Of course, one can always regard tensor fields on a manifold M as equivariant differential forms on GL(M) with values in an appropriate tensor product of copies of $\mathbb{R}^n$.

[19] In general, one refers to the tensor field $L_\mathbf{v} g$ as the *rate of* (*metric*) *strain* tensor for **v**.

[20] More generally, the characteristic foliation of any *k*-form is the foliation defined by the characteristic system of the ideal it generates.

$$\mu: M \to \mathfrak{G}^*,$$

such that for every $\mathfrak{g} \in \mathfrak{G}$:

$$i_{\hat{\mathfrak{g}}}\Omega = -d\mu(\mathfrak{g}).$$

Here, $\hat{\mathfrak{g}}$ refers to the fundamental vector field on M that is associated with $\mathfrak{g}$:

$$\hat{\mathfrak{g}}(x) = \frac{d}{ds}\bigg|_{s=0} \exp(\mathfrak{g}s)x.$$

Some sufficient conditions for a canonical group action on M to admit a momentum map are:

    *i)*    M is Haussdorff and simply connected.
    *ii)*    M is symplectic and $\mathfrak{G} = [\mathfrak{G}, \mathfrak{G}]$.

When M is symplectic, the momentum map satisfies the condition:

$$\hat{\mathfrak{g}} = \nabla_\Omega \mu(\mathfrak{g}).$$

If the (pre)-symplectic structure is exact $\Omega = d\theta$ such as with $T^*(M)$, and the action of G preserves $q$ then the action of G is also canonical and admits a momentum map with the property that:

$$\mu(\mathfrak{g}) = \theta(\hat{\mathfrak{g}}).$$

To relate this to variational problems, note that if the action of G on $J^1(\mathbb{R}, M)$ preserves the Lagrangian 1-form $\mathcal{L}d\tau$ then it also acts on $J^1(M, \mathbb{R})$ and preserves the corresponding Poincaré-Cartan form. Hence, the latter action is canonical and admits a momentum map. We can rephrase the classical theorem as:

**Nöther's theorem**:
> *If a Lie group* G *acts canonically on a (pre)-symplectic manifold* (M, $\Omega$) *that admits a momentum map* $\mu$ *then* $\mu$ *is constant on each leaf of the characteristic foliation of* $\Omega$.

In the case where G is the group of translations the value of $\mu$ represents linear momentum and when G = SO(3), it would represent angular momentum.

B. *Wave motion.* The two aspects of the problem of finding geometrical and topological invariants of wave motion that we will examine for the time being are concerned with the contact structure that is associated with wave motion and the aforementioned Godbillon-Vey form.

It has been observed [**4**] that wave motion implies a contact structure on space since one is not concerned with merely the motion of the points of space $\Sigma$ but with the motion of hyperplanes in their tangent spaces; one could regard this as simply a statement of the fact that wave motion is generally governed by second-order PDE's. The hyperplanes in question are tangent to the isophases, hence, orthogonal to the propagation vector. Since these hyperplanes are annihilated by the propagation 1-form $k$ that is metric-dual to that vector, as well as any other 1-forms in the 1-dimensional subspace of $T^*(\Sigma)$ that it spans, one can regard the association of a tangent hyperplane to each point $x \in \Sigma$ as equivalent to the association of a line through the origin of $T^*_x(\Sigma)$. Hence, one can define [2] a contact structure on $\Sigma$ by a section of PT*($\Sigma$), the projective cotangent bundle to $\Sigma$.

To the extent that wave motion should map the hyperplane at one point to the hyperplane at another point, one should expect it to proceed by way of *contactomorphisms*, i.e., diffeomorphisms of $\Sigma$ that preserve the chosen contact structure. Of course, if we wish to go beyond diffeomorphisms then we will expect to rethink the contact structure as well. However, we shall defer that to a later work.

One of the challenges to physics is to give a more physically intuitive characterization of the Godbillon-Vey form that is defined by a codimension-one foliation. Hopefully, if this present study has accomplished anything, it would be to emphasize the role of codimension-one foliations in the context of wave motion. Apropos of that, we make some observations about the construction of the Godbillon-Vey form for such a foliation of a proper time simultaneity leaf $\Sigma$.

Although the 1-form $\theta$ that generates the foliation is question mostly seems to define the shape of the wave fronts, it also seems to represent a sort of covelocity 1-form associated with the motion of the points of wave fronts. In that light, $d\theta$ would represent the vorticity of the flow of such points, which is a type of angular velocity. To express the vorticity as $d\theta = \eta \wedge \theta$ suggests that $\eta$ is essentially a "radius 1-form," that makes the vorticity into an orbital angular velocity for the flow defined by $\theta$ the same way that $r \wedge u$ defines the orbital angular velocity of the covelocity, relative to $r$. However, unlike the local radius 1-form $r = x_i dx^i$ that is defined by a local coordinate chart, which is closed – i.e., irrotational – $\eta$ might have non-vanishing vorticity, which is in the plane spanned by $\eta$ and $\theta$. Hence, one might regard the Godbillon-Vey form as the Frobenius form that is associated with the integrability of a radius 1-form.

VI. DISCUSSION. So far, the only kind of cobordism we have considered is the cylindrical kind $\Sigma \times I$ because we have not allowed our motion to have singular points. This has ruled out the possibility of topology-changing processes, and, in so doing, we have really not needed the extra generality of foliated cobordism. Up to this point, we could just as well have considered only concordance of space foliations.

In order to take the full advantage of the notion of foliated cobordism we must consider the role of singularities in more detail. In the case of particle motion, this would involve considering vector fields with zeroes – i.e., flows with fixed points. In the case of wave motion, the singularities are more involved. The most common types that are discussed in traditional continuum mechanics are sources and discontinuities, and caustics; these singularities also play an important role in optics. Because the methods and issues that are associated with singular foliations and wave singularities are beyond the scope of the present article, we shall return to them in a sequel work.

Most of the mathematical methods that were touched upon in the present article could be expanded upon considerably, such as the deeper nature of the role of foliated cobordism, the geometrical and topological nature of SO(2)-structures on spacetime, and the more general mathematical theory of wave equations. Of course, if the objective is to illuminate the deeper nature of the motion of physical waves, one must be wary of the ever-present temptation to degenerate into unbounded mathematicizing; after all, one is dealing with an infinite-dimensional manifold of possibilities. Perhaps theoretical physicists should, like Ulysses, lash themselves to the helm and pour wax in the ears of their research assistants, so as not to be lured onto the rocks of their doom by the seductive sirens of pure mathematics.

One of the most definitive aspects of wave motion in the eyes of quantum physics is the manner by which waves from different sources combine and "interfere," so it stands to reason that this process should also be incorporated into the context of foliations. One can immediately discern that a major modification has to be made since the interference patterns produced by even the simplest case of two distinct point sources can be more complex than the situations that were discussed in the present analysis.

Since the stated objective of the present article was to move in the direction of a "geometrical interpretation" for the wave mechanics of quantum physics, it would be worthwhile to sketch out a program for the development of "geometrical wave mechanics." The first step is to resurrect the long-discredited methods of continuum mechanics as a branch of physics. In particular, one should formulate them in terms of Lorentz manifolds and their various G-structures; the early work of the Cosserat brothers made a respectable start in that direction, although their work was still non-relativistic. Moreover, one should keep in mind the insights that physics has acquired from condensed matter concerning the structure of the vacuum state. From there, one would want to formulate the motion of waves in media that are modeled by such structures on spacetime. Hopefully, by the time one has examined the nature of wave singularities of the various types, some intuition concerning that way that one accounts for the motion of massive and massless waves as it appears in the atomic and sub-atomic domain might emerge.

## ACKNOWLEDGEMENTS

The author would like to thank the University of the Ozarks for providing a congenial environment in which to do research, Yuriy Krasnov for numerous discussions concerning the subject of the true place of topology in physical modeling, and Rafael Sorkin for first introducing the author's attention to the notion of topology-changing processes and Lorentz cobordism.